\documentclass[twocolumn]{aastex63}

\usepackage{amsmath}
\usepackage{amssymb}
\usepackage{textcomp}
\usepackage{comment}
\usepackage{multirow}

\newcommand{\sr}{${\rm Sr}$}
\newcommand{\he}{${\rm He}$}
\newcommand{\h}{${\rm H}$}
\newcommand{\rprocess}{\textit{r}-process}
\newcommand{\sprocess}{\textit{s}-process}

\newcommand{\refsec}[1]{{Sec.~(\ref{#1})}}
\newcommand{\reffig}[1]{{Fig.~(\ref{#1})}}
\newcommand{\refeq}[1]{{Eq.~(\ref{#1})}}
\newcommand{\reftab}[1]{{Table~\ref{#1}}}
\newcommand{\refapp}[1]{{Appendix~\ref{#1}}}

\submitjournal{ApJ}

\shorttitle{Very light elements in BNS mergers}
\shortauthors{Perego et al.}

\begin{document}

\title{Production of very light elements and strontium in the early ejecta of neutron star mergers}

\correspondingauthor{Albino Perego}
\email{albino.perego@unitn.it}

\author[0000-0002-0936-8237]{Albino Perego}
\affiliation{Dipartimento di Fisica, Università di Trento, Via Sommarive 14, 38123 Trento, Italy}
\affiliation{INFN-TIFPA, Trento Institute for Fundamental Physics and Applications, ViaSommarive 14, I-38123 Trento, Italy}

\author[0000-0003-0309-4666]{Diego Vescovi}
\affiliation{Gran Sasso Science Institute, Viale Francesco Crispi, 7, 67100 L'Aquila, Italy}
\affiliation{Istituto Nazionale di Fisica Nucleare, Sezione di Perugia, Via A. Pascoli snc, 06123 Perugia, Italy}
\affiliation{Istituto Nazionale di Astrofisica, Osservatorio d'Abruzzo, Via Mentore Maggini snc, 64100 Teramo, Italy}

\author{Achille Fiore}
\affiliation{INAF Osservatorio Astronomico di Padova, Vicolo dell’Osservatotio 5, I-35122, Padova, Italy}
\affiliation{Department of Physics and Astronomy Galileo Galilei, University of Padova, Vicolo dell'Osservatorio 3, I-35122, Padova, Italy}

\author{Leonardo Chiesa}
\affiliation{Dipartimento di Fisica, Università di Trento, Via Sommarive 14, 38123 Trento, Italy}

\author[0000-0002-7941-5692]{Christian Vogl}
\affiliation{Max-Planck-Institut f\"ur Astrophysik, Karl-Schwarzschild-Str. 1, 85748 Garching, Germany}
\affiliation{Exzellenzcluster ORIGINS, Boltzmannstr. 2, 85748 Garching, Germany}

\author{Stefano Benetti}
\affiliation{INAF Osservatorio Astronomico di Padova, Vicolo dell’Osservatotio 5, I-35122, Padova, Italy}

\author{Sebastiano Bernuzzi}
\affiliation{Theoretisch-Physikalisches Institut, Friedrich-Schiller Universit\"{a}t Jena, 07743, Jena, Germany}

\author{Marica Branchesi}
\affiliation{Gran Sasso Science Institute, Viale Francesco Crispi, 7, 67100 L'Aquila, Italy}
\affiliation{INFN - Laboratori Nazionali del Gran Sasso, I-67100, L`Aquila (AQ), Italy}

\author{Enrico Cappellaro}
\affiliation{INAF Osservatorio Astronomico di Padova, Vicolo dell’Osservatotio 5, I-35122, Padova, Italy}

\author{Sergio Cristallo}
\affiliation{Istituto Nazionale di Astrofisica, Osservatorio d'Abruzzo, Via Mentore Maggini snc, 64100 Teramo, Italy}
\affiliation{Istituto Nazionale di Fisica Nucleare, Sezione di Perugia, Via A. Pascoli snc, 06123 Perugia, Italy}

\author[0000-0003-2024-2819]{Andreas Fl\"ors}
\affiliation{GSI Helmholtzzentrum f\"ur Schwerionenforschung, Planckstra\ss e 1, 64291 Darmstadt, Germany}

\author[0000-0002-0479-7235]{Wolfgang E. Kerzendorf}
\affiliation{Department of Physics and Astronomy, Michigan State University, East Lansing, MI 48824, USA}
\affiliation{Department of Computational Mathematics, Science, and Engineering, Michigan State University, East Lansing, MI 48824, USA}

\author{David Radice}
\affiliation{Institute for Gravitation \& the Cosmos, The Pennsylvania State University, University Park, PA 16802} 
\affiliation{Department of Physics, The Pennsylvania State University, University Park, PA 16802}
\affiliation{Department of Astronomy \& Astrophysics, The Pennsylvania State University,University Park, PA 16802}

\date{\today}
\begin{abstract}
We study the production of very light elements ($Z < 20$) in the dynamical and spiral-wave wind ejecta of binary neutron star mergers by combining detailed nucleosynthesis calculations with the outcome of numerical relativity merger simulations. All our models are targeted to GW170817 and include neutrino radiation. We explore different finite-temperature, composition dependent nuclear equations of state and binary mass ratios, and find that hydrogen and helium are the most abundant light elements.
For both elements, the decay of free neutrons is the driving nuclear reaction. In particular, $\sim 0.5-2 \times 10^{-6} M_{\odot}$ of hydrogen are produced in the fast expanding tail of the dynamical ejecta, while $\sim  1.5-11 \times 10^{-6} M_{\odot}$ of Helium are synthesized in the bulk of the dynamical ejecta, usually in association with heavy \rprocess{} elements.
By computing synthetic spectra, we find that the possibility of detecting hydrogen and helium features in kilonova spectra is very unlikely for fiducial masses and luminosities, even when including non local thermodynamics equilibrium effects. The latter could be crucial to observe He lines a few days after merger for faint kilonovae or for luminous kilonovae ejecting large masses of helium.
Finally, we compute the amount of strontium synthesized in the dynamical and spiral-wave wind ejecta, and find that it is consistent with (or even larger than, in the case of a long lived remnant) the one required to explain early spectral features in the kilonova of GW170817.
\end{abstract}

\keywords{Neutron stars -- Explosive nucleosynthesis -- R-process}

\section{Introduction}\label{sec:intro}

Binary neutron star (BNS) mergers are primary sites for the production of heavy elements in the Universe through the rapid neutron capture process \citep[\rprocess, e.g.][]{Symbalisty:1982a,Eichler:1989ve,Freiburghaus:1999}. 
This association  was confirmed by the detection of the kilonova AT2017gfo
\citep{Arcavi:2017xiz,Coulter:2017wya,Drout:2017ijr,Evans:2017mmy,Kasliwal:2017ngb,Nicholl:2017ahq,pian17,Smartt:2017fuw,Soares-santos:2017lru,Tanvir:2017pws}
as electromagnetic (EM) counterpart of the BNS gravitational wave (GW) signal GW170817 \citep{Abbott2017,Abbott2017b}.
The luminosity evolution of the UV/visible/IR transient AT2017gfo is indeed in agreement with the heating rate and opacity expected from a distribution of freshly synthesized \rprocess~ elements \citep[e.g.][]{Villar.etal:2017,kas17,Tanaka:2017qxj,Wollaeger:2017ahm,Perego:2017wtu}.

A few days after merger the spectrum of AT2017gfo reveals emission and absorption features qualitatively compatible with the forest of lines expected for matter rich in heavy elements (in particular, lanthanides and actinides). However, the firm identification of spectral features attributable to specific elements whose mass number is larger than $A \sim 100$ was so far not robust \citep[see however][for interesting attempts]{Smartt:2017fuw,Gillanders.etal:2021}.
The main difficulties here lie in the huge number of possible bound-bound and bound-free transitions that provide the bulk of the photon opacity in matter enriched in heavy elements, and in our still poor knowledge of these atomic transitions.
Additionally, the high expansion speed of the matter expelled by BNS mergers (ejecta) and its non uniform spatial distribution are expected to cause a significant and non-trivial line broadening.
However, in the early kilonova phases most of the ejecta are still opaque to radiation: only the fastest ejecta have become transparent and form an atmosphere that can alter the thermal emission coming from the underlying photosphere. The composition of this thin atmosphere could actually provide spectral features whose origin is possibly easier to identify. Indeed, the spectrum at 1.5 days of AT2017gfo resembles a black body emission with a significal residual around $8000$ \AA{} whose analysis suggested the presence of strontium \citep[Sr,][]{watson19}, a light \rprocess~element whose production in the Universe is however dominated by the slow neutron capture \citep[see e.g.][]{pra2020}.

The discovery of EM counterparts of BNS mergers detected in GWs is a challenge that often requires prolonged observations of multiple candidate transients, due to the uncertainties in the sky localization of the source. The identification of lines in their spectra and the comparison with the expected abundances can help discriminate between more and less plausible candidates.
This approach heavily relies on detailed theoretical modelling of the ejecta from BNS mergers and of the subsequent early kilonova emission.
Different mechanisms, acting on different timescales, 
are responsible for multiple ejecta components, whose properties mainly depend on the still uncertain equation of state (EOS) of nuclear matter and on the binary mass ratio \citep[see e.g.][and references therein]{Radice:2020ddv}. 
The dynamical \citep[see e.g.][]{koro12,Bauswein:2013yna,Sekiguchi:2015dma,
palenzuela:2015dqa,radice:2016dwd,Lehner:2016lxy,sekiguchi:2016bjd,
Foucart.etal:2016, bovard:2017mvn,Radice:2018pdn} and spiral-wave wind \citep{ned19,ned20} ejecta are the earliest and fastest ejecta, thus becoming transparent within the very first days and possibly providing key spectral features. Larger amounts of matter are expelled later in the form of baryonic disk winds \citep[see e.g.][]{fer13,per14,Metzger:2014ila,sie14,Martin:2015hxa,just15,lip17b,Siegel:2017jug,fuj18,fer19,Miller:2019dpt}. In these cases the ejection mechanisms are, for example, turbulent viscosity of magnetic origin, neutrino irradiation, magnetic pressure. This larger amount of mass expanding with lower speed (compared to the dynamical ejecta) is expected to become transparent only after a few days. 

In this paper, we investigate in a systematic way the production of very light elements (lighter than calcium) in BNS mergers, focusing in particular on hydrogen and helium, based on detailed merger simulations. For the first time, we directly connect the thermodynamics conditions for their production to the binary properties (mass ratio and EOS), and we study their early detectability in kilonova spectra.
The presence of H in the dynamical ejecta of BNS mergers was already predicted by \citet{Metz14,just15}. Indeed, in these works it was noticed that the head of the dynamical ejecta can contain $\sim 10^{-4} M_{\odot}$ of free neutrons expanding at $\gtrsim 0.4 c$, not captured by seed nuclei due to their sudden drop in density. In addition to producing a peculiar neutron-powered precursors of kilonovae \citep{Metz14}, this fast expanding matter would provide an envelope of hydrogen around the ejecta. In this paper, we want to check the amount of \h{} with respect to the inclusion of weak interactions in merger simulations and in the modelling of the merger through full general relativistic simulations \citep[see also][]{Ishii.etal:2018,Manu.etal:2020}. Moreover, we want to test if \h{} can give possible spectral features.
Also the production of \he{} has been reported in the analysis of the abundances obtained in the dynamical ejecta of BNS mergers \citep[see e.g.][]{wanajo2014}. However, its origin and its dependence on the binary parameters and on the EOS have never been investigated, as well as its spectral detectability.

In addition to the very light elements, we also study the production of strontium since this element was claimed to be detected in AT2017gfo. We want to test if its inferred amount is compatible with our predictions and if this information can help discriminate between different merger models of GW170817.

The paper is structured as follows. In \refsec{sec:method}, we present the methods used in our analysis: in particular, the BNS merger simulations, \refsec{subsec:simulations}; the nucleosynthesis calculations, \refsec{subsec:nucleosynthesis}; and the kilonova spectrum model, \refsec{subsec:kilonova spectrum}.
Our results are presented in \refsec{sec:results}, focusing first on the nucleosynthetic yields, \refsec{subsec:results nucleo}; then on the nuclear processes responsible for their production, \refsec{subsec:nucleosynthesis analysis}, and on the analysis of the ejecta conditions, \refsec{subsec:results ejecta}; finally on the kilonova spectral features, \refsec{subsec:results kilo spectra}. We summarize and discuss our results in \refsec{sec:discussion} and \refsec{sec:conclu}.

\section{Method}\label{sec:method}

\begin{table*}
    \centering
    \begin{tabular}{c|c|c|c|c|c|c|c}
         \hline\hline
         \multicolumn{8}{c}{Dynamical ejecta}  \\ \hline
         Model	& M1 & M2 &	
         Viscosity \& Resolution &
         $m_{\rm ej}$ & $m_{\rm H}$	& $m_{\rm He}$	& $m_{\rm Sr}$ \\ {~}	& $[M_{\odot}]$ & $[M_{\odot}]$ &	
         {~} &
         $[10^{-3}M_{\odot}]$ & $[10^{-6}M_{\odot}]$ & $[10^{-6}M_{\odot}]$ & $[10^{-5}M_{\odot}]$ \\
         \hline
         BLh\_equal & 1.364 & 1.364 & vis: (LR,SR); no-vis: (LR,SR,HR) & $1.37^{+0.29}_{-0.25}$ & $1.52^{+0.5}_{-0.98}$ & $3.87^{+4.51}_{-2.21}$ & $3.01^{+0.47}_{-0.53}$ \\ \hline
         BLh\_unequal & 1.856 & 1.020 & vis: (SR,HR); no-vis: (SR,HR) &  $9.2^{+0.14}_{-0.15}$ & $0.78^{+0.29}_{-0.33}$ & $9.25^{+1.52}_{-1.84}$ & $0.25^{+0.14}_{-0.14}$ \\ \hline
         DD2\_equal & 1.364 & 1.364 & vis: (LR,SR,HR); no-vis: (LR,SR,HR) & $1.36^{+0.78}_{-0.13}$ & $1.70^{+0.19}_{-0.37}$ & $3.12^{+1.99}_{-1.43}$ & $2.91^{+1.85}_{-2.34}$ \\ \hline
         \multicolumn{8}{c}{~}  \\
    \end{tabular}
        \begin{tabular}{c|c|c|c|c|c|c|c}
         \hline\hline
         \multicolumn{8}{c}{Spiral wave wind ejecta}  \\ \hline
         Model	& M1 & M2 &	
         Viscosity \& Resolution & 
         $R_{\rm wind}$ & $X_{\rm H}$	& $X_{\rm He}$ & $X_{\rm Sr}$ \\ 
         {~}	& $[M_{\odot}]$ & $[M_{\odot}]$ &	
         {~} 
         & $[10^{-1} M_{\odot}~{\rm s}^{-1}]$ &
         $[10^{-8}]$ & $[10^{-5}]$ & $[10^{-2}]$ \\
         \hline
         BLh\_equal & 1.364 & 1.364 & vis: (LR,SR); no-vis: (LR,SR) & $1.57 \pm 0.42$ 
         & $0.92^{+0.40}_{-0.33}$ & $9.80^{+10.12}_{-8.28}$ & $2.12^{+0.31}_{-0.49}$ \\ \hline 
         DD2\_equal & 1.364 & 1.364 & vis: (LR,SR,HR); no-vis: (LR,SR) & $1.58\pm 0.04$ 
         & $6.64^{+2.49}_{-2.07}$ & $4.57^{+2.34}_{-1.20}$ & $3.58^{+0.65}_{-0.57}$ \\ \hline 
    \end{tabular}
    \caption{Properties of the dynamical ejecta (top) and of the spiral wave wind ejecta (bottom). For each BNS merger model we indicate the masses of the two NSs ($M_1$ and $M_2$) and the corresponding set of available simulations, differing by resolution and physical viscosity of turbulent origin. For the dynamical ejecta, we report the total, \h{}, \he{} and \sr{} ejected masses. For the spiral wave wind ejecta, we report the average ejection rate and the \h{}, \he{} and \sr{} mass fractions. The reported numbers are the mean values, averaged over the set of simulations, while the errors corresponds to the distance between the average and the largest or smallest values.}
    \label{tab:summary dynamical}    
\end{table*}

\subsection{Binary neutron star merger simulations} \label{subsec:simulations}
Nucleosynthesis in BNS mergers 
depends mainly on three physical parameters: the specific entropy ($s$), the electron fraction ($Y_e$), and the expansion timescale ($\tau$) \citep{hoff97}. 
BNS merger ejecta 
cannot be characterized by a single value of these parameters. A distribution in the $(s,Y_e,\tau)$ space is instead expected.

In this work, we consider results of Numerical Relativity (NR) simulations performed with the \texttt{WhiskyTHC} code \citep{Radice:2012cu,rad14,rad14b}.
The latter is a NR code that solves the Einstein’s  equations in the 3+1 Z4c free-evolution scheme \citep{Bernuzzi:2009ex,Hilditch:2012fp} coupled to general relativistic hydrodynamics on adaptive mesh-refinement grids. \texttt{WhiskyTHC} employs high-resolution shock capturing algorithms
and implements finite-temperature, composition dependent nuclear EOSs, an approximate neutrino transport scheme \citep{radice:2016dwd,Radice:2018pdn},  and the general-relativistic large eddy simulations method (GRLES) for turbulent viscosity of magnetic origin \citep{Radice:2017zta,Radice:2020ids}. The code was specifically designed to model the late inspiral, merger and post-merger phase of BNS mergers.

We consider 3 models whose chirp mass is targeted to the GW170817 event \citep{Abbott2017} labelled as BLh\_equal, BLh\_unequal, and DD2\_equal. 
The first and the third one have $M_1 = M_2 = 1.364 M_{\odot}$, while the second one has $M_1 = 1.856 M_{\odot} > M_2 = 1.020 M_{\odot}$. 
The BLh\_equal and BLh\_unequal models employ the softer BLh nuclear EOS \citep{Logoteta:2020}, 
an hadronic EOS whose high density part has been derived using the finite temperature extension of the Brueckner-Bethe-Goldstone quantum many-body theory in the Brueckner-Hartree-Fock approximation. This EOS predicts a maximum mass of $2.10 M_{\odot}$ for a cold, non-rotating NS. The DD2\_equal model uses the stiffer HS(DD2) EOS \citep{Typel:2009sy,Hempel:2011mk}.
This EOS was derived in the framework of relativistic mean field models, uses density-dependent couplings at high density, and predicts a maximum NS mass of $ 2.42M_{\odot}$. Both these EOSs are consistent with current nuclear and astrophysical constraints and roughly bracket uncertainties in the properties of matter above nuclear densities.

For all simulations the adaptive mesh refinement is characterised by seven nested grids with 2:1 refinement level. Each physical set-up is run at least at two different resolutions, sometimes even at three.
The linear resolution in the finest level is of $\sim 246{\rm m}$, $\sim 187{\rm m}$, $\sim 125{\rm m}$ for the low, standard, and high resolution case, respectively. We denote the three cases as LR, SR and HR, respectively.
All models include neutrino radiation and, in particular, neutrino absorption in optically thin conditions. The latter is crucial to correctly predict the composition of the dynamical ejecta \citep{wanajo2014,Radice:2018pdn,Foucart.etal:2016}.
For each model we consider both simulations with and without turbulent viscosity.
All the simulations employed in this work were presented in \citet{per19,ned19,bern20,ned20} where more details can be found.

Tidal torques and shock waves produced by the bouncing remnant unbind matter within a few milliseconds (the so-called dynamical ejecta). For equal mass binaries, a softer EOS produces stronger shocks and larger shock-heated ejecta, while tidal ejection within a crescent across the equatorial characterises very unequal mass mergers. 
The dynamical ejecta are obtained by applying the geodetic extraction criterion.
A summary of the amount of dynamical ejecta for the three different models, as well as a list of the employed simulations, can be seen in the upper part of \reftab{tab:summary dynamical}.
The mass values are computed as the arithmetic average of the distribution of the ejecta masses obtained by considering the different available resolutions and by including both viscous and unviscous simulations. The (possibly asymmetric) errors correspond to the minimum and the maximum differences between the average and distribution of the actual masses.
Additionally, the DD2\_equal and BLh\_equal simulations were extended up to several tens of milliseconds after merger (and in particular, up to $\sim$90-100 ms post merger in the SR cases) showing the development of a $m=1$ spiral arm in the central remnant. The spiral arm propagates into the disk, transporting angular momentum outwards and producing a matter outflow in the form of a spiral wave wind.
The presence of this wind is directly related to the presence of a non-collapsed remnant in the center (this justifies why we did not compute the wind for the BLh\_unequal model, in which a prompt BH formation occurs). Moreover the corresponding ejection rate did not show signs of attenuation at the end of the simulations.
We decided to estimate the ejection rate in the form of a spiral wave wind by considering the longest simulations available, i.e. the SR ones. The SR simulations of the BLh\_equal model produce $1.08 \times 10^{-2} M_{\odot}$ around 105 ms for the simulation without viscosity, and $1.61 \times 10^{-2} M_{\odot}$ around 90 ms for the viscous simulation of spiral wave wind ejecta, obtained by using the Bernoulli extraction criterion.
The SR simulations of the DD2\_equal model produce $1.23 \times 10^{-2} M_{\odot}$ around 80 ms post merger for the simulation without viscosity, and $1.58 \times 10^{-2} M_{\odot}$ around 107 ms post merger for the viscous simulation.
When considering that this wind develops starting $10{\rm ms}$ after merger, we estimate the spiral wave wind ejecta rate, $R_{\rm wind}$, to be $(1.57 \pm 0.42) \times 10^{-1} M_{\odot}~{\rm s^{-1}}$ and $(1.58 \pm 0.04) \times 10^{-1} M_{\odot}~{\rm s^{-1}}$
for the BLh\_equal and DD2\_equal model, respectively (see the bottom part of \reftab{tab:summary dynamical}).
Once again, for each model the actual values and their uncertainties have been computed using the average and the differences between the two available simulations.

\subsection{Nucleosynthesis}
\label{subsec:nucleosynthesis}
To compute time-dependent yield abundances we use the publicly available nuclear network \texttt{SkyNet} including 7843 isotopes up to ${}^{337}{\rm Cn}$ \citep{lipp17}. 
We employ the latest JINA REACLIB database \citep{cybu10}, while using the same set-up as in ~\citet{lipp15} for the other input nuclear physics. In particular, strong inverse rates are computed assuming detailed balance. Spontaneous and neutron-induced fission rates are taken from \citet{fran47,pano10}, adopting fission barriers and fission fragment distributions from \citet{mamd01,wahl02}.
The default version of \texttt{SkyNet} does not include $\beta$-delayed fission reactions. However, we do not expect them to affect the synthesis of light elements, but only of neutron-rich heavy nuclei and therefore the final abundances near the second $r$-process peak (see, e.g., \citealt{mumpower18}). Also, they do not significantly alter the abundances of nuclei surviving to fission after neutron freeze out and undergoing $\alpha$-decay, and so we also assume that their contribution to \he{} abundance in low Y$_e$ conditions is negligible.
Nuclear masses are taken from the REACLIB database, which includes experimental values where available and theoretical masses from the finite-range droplet macroscopic model \citep[FRDM,][]{moll16} otherwise.

\texttt{SkyNet} requires time-dependent trajectories of Lagrangian fluid elements to predict the temporal evolution of the abundances. 
We initialised all trajectories in NSE at $T_0 = 6~{\rm GK}$. For a given electron fraction and entropy, the NSE solver determines the corresponding initial density, $\rho_0$, by considering a fully ionised ideal gas of ions, electrons and photons.
After that, matter density evolves first through an
exponential phase and then to a homologous expansion \citep{lipp15}:
\begin{equation}
    \rho(t) = 
    \begin{cases}
    \rho_{0} e^{-t / \tau} & {\rm for~}t \leqslant 3 \tau, \\
    \rho_{0}\left( 3 \tau /\left( e t\right) \right)^{3}  & {\rm otherwise.}
    \end{cases}
    \label{eq:density expansion}
\end{equation}
The tracer temperature is evolved consistently to the expansion, accounting for nuclear heating. The tracer and the abundances of all relevant nuclear species are evolved up to $10^9~{\rm s}$.

To cover the relevant intervals we perform extensive nucleosynthesis calculations over wide ranges of $\tau$, $Y_e$, and $s$, namely: $0.5 \leq \tau~{\rm [ms]} \leq 200$, $1.5 \leq s~[k_{\rm B}~{\rm baryon^{-1}}] \leq 300$, and  $0.01 \leq Y_e \leq 0.48$, by constructing a $18 \times 26 \times 25$ regular grid (approximately logarithmic in the two former quantities and linear in the latter). The above ranges span the relevant expected intervals for the ejecta properties of compact binary mergers.

We finally obtain time dependent abundances from the convolution of the yields tabulated with \texttt{SkyNet} 
with the distribution of ejecta properties from the simulations.
From each BNS simulations we extract the mass weighted distribution of the ejecta in the $(s,Y_e,v_{\infty})$ space by inspecting the properties of the unbound matter on a coordinate sphere of radius 294km. The solid angle is discretized in $N_\theta=51$ polar and $N_\phi=51$ azimuthal angular bins, uniform in both the $\theta$ and $\phi$ angles. We stress that we do not perform any averaging procedure over the angular variables.
$v_{\infty}$ is the asymptotic velocity computed as 
$v_{\infty} = c \left( 1-\gamma^{-2} \right)^{1/2}$.
In the case of the geodetic criterion, $ \gamma = u_t$ is the time component of the 4-velocity, while for the Bernoulli criterion, $\gamma = u_t h$, where $h$ is the relativistic specific enthalpy per baryon.
The expansion timescale is computed starting from the density and velocity obtained from the simulations at the extraction radius, following the procedure introduced in \citet{radice:2016dwd, Radice:2018pdn}.


\subsection{Kilonova spectra}
\label{subsec:kilonova spectrum}

The possible presence of light elements 
in BNS merger ejecta raises the question of whether they can produce recognisable features in the observable spectra. To address this issue, we use the open-source spectral synthesis code \texttt{TARDIS} \citep{tardis} to produce spectral models for the predicted ejecta abundances and physical conditions. \texttt{TARDIS} is a Monte Carlo radiative-transfer code that operates in one dimension assuming the ejecta are spherically symmetric. 
In particular, it prescribes a thermal emission at the photosphere and then predicts the spectrum emerging after the radiation has interacted with the above atmosphere at a certain time after merger.
Matter within the atmosphere is discretized by Lagrangian radial mass shells of monotonically increasing speed in velocity space. The code inputs include the luminosity at the photosphere, the density profile and the composition of the ejecta between the photosphere and the head of the ejecta.
A set of atomic data for the relevant ions and the adopted approximations in computing the ionization/excitation state of the atoms in the ejecta, as well as the handling of matter-radiation interactions, are also required \citep[see references in][]{tardis,boyle17,vogl19}. 

The presence of lanthanides and actinides significantly increases matter opacity in BNS merger ejecta \citep{kas13,TanakaHotokezaka:2013} such that radiation drives the expanding ejecta toward 
local thermodynamical equilibrium
(LTE) conditions \citep{kas17}.
However we cannot exclude the possibility that the \h{} and \he{} lines are boosted by non local thermodynamical equilibrium (NLTE) effects similar to those observed in the ejecta of supernovae (SNe). 
Indeed, helium excitation and ionisation rates in supernova ejecta
are strongly affected by non-thermal electron collisions, produced by $\gamma$-rays resulting from the radioactive decay of ${}^{56}{\rm Ni}$ \citep{Graham:1988,Lucy:1991,Mazzali.etal:1998,Hachinger.etal:2012}.
The significant sensitivity of \he{} to these effects is due to the large energy gap (20 eV) between the He I ground state and its first excited state, and to the metastability of its first two excited states.
Since \he{} in BNS merger ejecta is often embedded in radioactive material and the $Q$-value of ${}^{56}{\rm Ni}$ decay is comparable to the one of $r$-process element decays during the kilonova timescale, it is reasonable to assume that similar effects could occur also in kilonovae.
\texttt{TARDIS} models LTE conditions between matter and radiation, but it also offers a variety of approximate NLTE treatments of ionization and excitation (for example the nebular approximation for the ionization balance).
Among them, \texttt{TARDIS} includes an analytical approximation for the NLTE helium level population that was developed for the \he{}-rich ejecta of double detonation type Ia supernovae \citep{boyle17}. 

Additionally, strong departures from LTE can arise even without the presence of high-velocity electrons. The radiation field above the photosphere is typically dilute compared to that of a blackbody. In these conditions, the rate of collisions with thermal electrons is often too low to establish LTE populations for the low densities in the outer layers of the ejecta. The resulting NLTE effects significantly affect the hydrogen line strengths, as observed in type II SNe \citep{Takeda90,Takeda91,Duschinger95}.
To model these effects, we use the \texttt{TARDIS} version presented in \citet{vogl19}, which has already been applied to modeling H-rich ejecta in NLTE conditions \citep{Vogl2020}. This version of the code includes a more complete treatment of radiation-matter interactions including bound-free and free-free processes, as well as collisions between ions and thermal electrons. The modified code solves the statistical equilibrium equations for the ion and level number densities without relying on the Boltzmann and Saha equations.

In addition to the nucleosynthesis yields provided by our BNS and nucleosynthesis calculations, we require a model for the ejecta profile and photosphere evolution. In accordance with the symmetry employed by \texttt{TARDIS}, we consider the analytic spherically symmetric model presented in \citet{Wollaeger:2017ahm}. In particular, we assume a homologously expanding layer of ejecta of total mass $M$, average expansion speed $v_{\rm avg}$, and uniform gray opacity $\kappa$. 
The latter should be intended as an effective, average opacity related to
the more physical and detailed energy-dependent one and to the relevant radiation spectrum. Typical values of $\kappa$ for kilonovae range between 1 and a few tens ${\rm cm^2~g^{-1}}$, depending on the matter composition \citep[see e.g.][]{Tanaka.etal:2020}.
The density profile is described by:
\begin{equation}
    \rho(t,r) = \rho_0 \left( \frac{t}{t_0} \right)^{-3} \left( 1 - \frac{r^2}{\left( v_{\rm max} t \right)^2} \right)^3 \, ,
    \label{eq: density homologous expansion}
\end{equation}
where $v_{\rm max} = 128 v_{\rm avg} /63$ and $\rho_0 t_0^3 = 315 M/(64 \pi v_{\rm max}^3)$. The time dependent photospheric radius, $R_{\rm ph}(t)$, is defined by the condition:
\begin{equation}
    \int_{R_{\rm ph}(t)}^{R_{\rm max}(t)} \: \rho(t,r) \kappa~{\rm d}r = \frac{2}{3} \, ,
    \label{eq: photosphere}
\end{equation}
while the mass outside the photosphere is computed as:
\begin{equation}
    M_{>R_{\rm ph}}(t) = 4 \pi \int_{R_{\rm ph}(t)}^{R_{\rm max}(t)} \: \rho(t,r) r^2 \, {\rm d}r \, .
    \label{eq: transparent mass}
\end{equation}
Each fluid element expands with constant radial speed $v$ and the relation between the (Lagrangian) velocity coordinate and the (Eulerian) radial coordinate is $v=rt$.

\begin{figure}[!t]
\centering
\includegraphics[width = \linewidth]{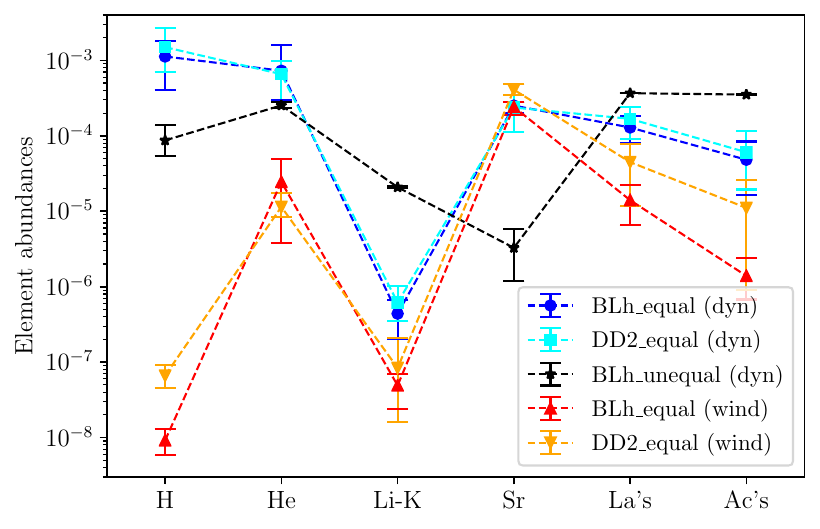} 
\caption{Number abundances of very light elements ($Z < 20$), strontium, lanthanides and actinides in the dynamical and spiral wave wind ejecta for the three BNS models considered in this work, at two days after merger. The abundances of lanthanides, actanides and of all elements between lithium and potassium are summed. \h{}, \he{} and \sr{} are robustly synthetized in the dynamical ejecta of equal mass mergers, while \sr{} is also significantly produced in the spiral wave wind ejecta. The production of elements between lithium and potassium is subdominant.}
\label{fig:abundances from NR simulations}
\end{figure}

\begin{figure*}[!t]
\centering
\includegraphics[width = \linewidth]{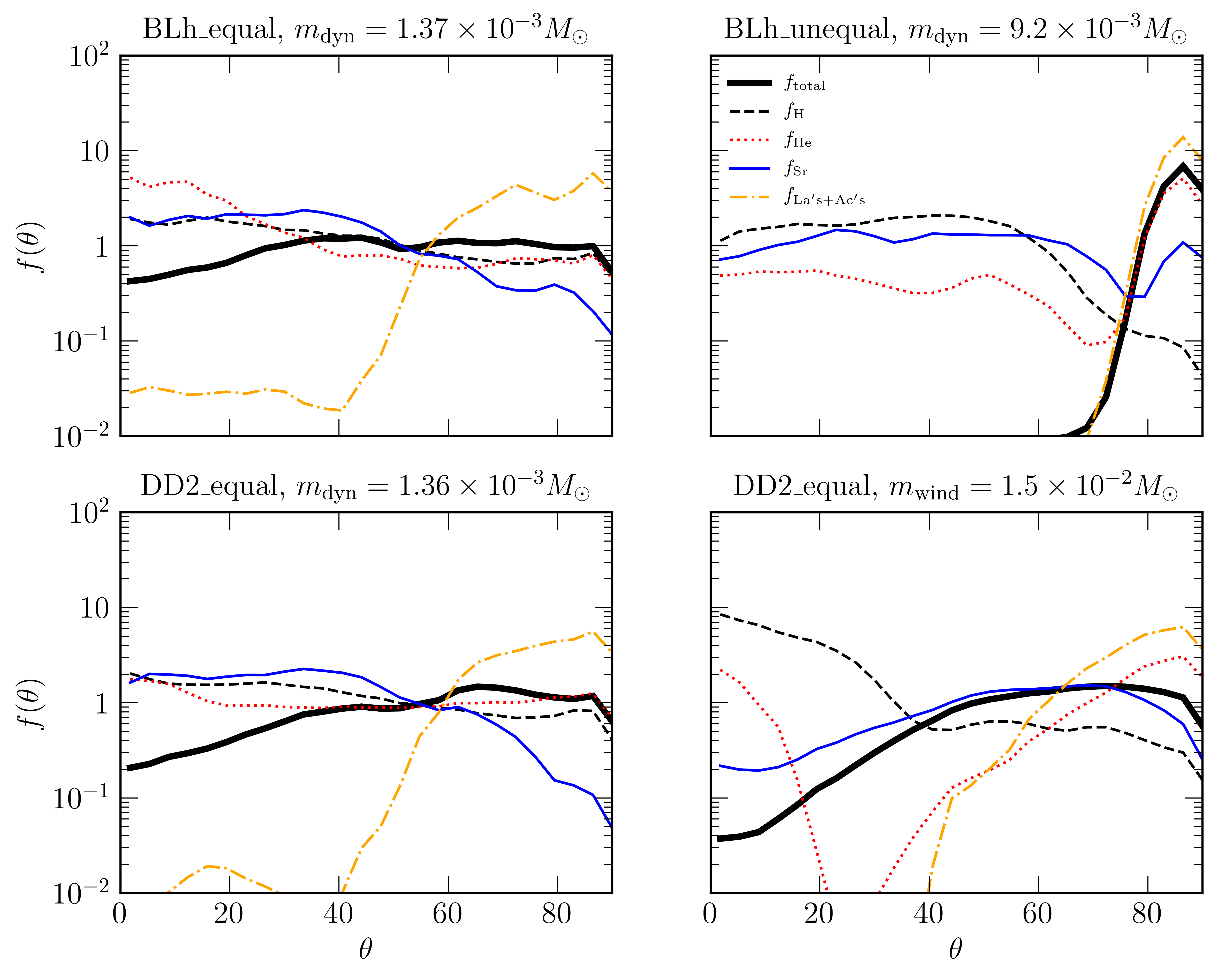} 
\caption{Polar distribution function (defined as the ratio between the actual angular distribution and the isotropic equivalent, averaged over the azimuthal angle) of the ejected mass for some of the elements reported in \reffig{fig:abundances from NR simulations} and \reftab{tab:summary dynamical}, as a function of the polar angle $\theta$. Distributions for which $f \sim 1$ are closer to an isotropic distribution. The distribution of the total ejecta (black and thick solid line) traces the actual presence of the ejecta. The four panels refer to the dynamical ejecta of the BLh\_equal (top-left), the dynamical ejecta of BLh\_unequal (top-right), the dynamical ejecta of DD2\_equal (bottom-left), and to the spiral wave wind ejecta of DD2\_equal (bottom-right). For each model, we considered unviscous, SR simulations as representative simulations.}
\label{fig:abundances angular distribution}
\end{figure*}

\section{Results}
\label{sec:results}

\subsection{Nucleosynthesis: overview}
\label{subsec:results nucleo}
In \reffig{fig:abundances from NR simulations} we present number abundances of selected elements in the ejecta of the three considered models, two days after merger. For each model, the abundance values are obtained as the averages over all available simulations, while the error bars are the maximum and minimum difference with respect to the average.
Among the lightest elements ($Z < 20$), \h~and \he~are the most abundant species, while all the elements between lithium and potassium are usually several orders of magnitudes less abundant ($Y \lesssim 10^{-5}$).
When considering the dynamical ejecta, the production of \h~and \he~ appears robust and their abundances vary only within one order of magnitude even when changing the EOS stiffness or the binary mass ratio.
\h{} and \he{} abundances are comparable to (or even larger than) lanthanides and actinides, as well as to Sr abundance, unless the binary is very asymmetric. Light element production is less significant in the spiral-wave wind ejecta, where the production of the first and second \rprocess~peak elements is favored. 
However, we stress that the spiral-wave wind ejecta (red and yellow line) should be always considered in combination with the dynamical ejecta coming from the same model. In the case of a long-lived remnant, due to the larger spiral-wave wind contribution, \h{} and \he{} tend to be slightly underproduced with respect to \sr{} and heavy \rprocess~elements.

In addition to the number abundances, our simulations can provide also the masses of the different elements. In the upper part of \reftab{tab:summary dynamical} we report the masses of \h{}, \he{} and \sr{} in the dynamical ejecta for the different merger models. Once again, the central values correspond to the arithmetic averages while the uncertainties to the largest difference between the average and the distribution of the actual values.
Since the spiral wave wind ejecta have not saturated by the end of our equal mass simulations and the precise amount of ejecta depends on the central remnant lifetime, in the bottom part of \reftab{tab:summary dynamical} we provide the mass fraction of the above elements in the spiral wave wind ejecta.

Matter ejection from BNS mergers is not isotropic and from our models we can extract the angular distributions for the different atomic species. In \reffig{fig:abundances angular distribution} we present the polar distribution factor for an element $i$, $f_i(\theta)$, defined as:
\begin{equation}
    f_i(\theta) = \frac{4 \pi}{M_i} \left( \frac{1}{ 2 \pi} \int_0^{2 \pi} \; \frac{{\rm d}M_i}{{\rm d}\Omega} \, {\rm d}\phi \right) \, ,
\end{equation}
i.e. the ratio between the actual angular distribution of the mass of a certain element (averaged over the azimuthal angle) and the equivalent isotropic one ($M_i/ 4\pi$). This factor measures how much the obtained ejecta distributions differ from an isotropic distribution. For the dynamical ejecta expelled by equal mass mergers (left panels), the total amount of mass (black solid line) shows a larger distribution at lower latitudes and a clear decrease (more pronounced in the case of the stiffer DD2 EOS) moving toward the poles. This is consistent with previous results \citep[see e.g.][]{Radice:2018pdn} and testifies that, despite being emitted over the entire solid angle, the dynamical ejecta are not isotropic. \h{} and \he{} are distributed at all latitudes but in associations with different elements: close to the equator, together with heavier elements (e.g. with lanthanides and actinides); at high latitude, with lighter $r$-process elements (as \sr{}). This suggests a distinct origin, especially for \he{}.
For the dynamical ejecta produced by very asymmetric mergers (top-right panel), most of the ejecta is concentrated across the equator \citep{Lehner:2016lxy,sekiguchi:2016bjd,bern20}, where heavy elements and \he{} are significantly synthesised, while \h{} and \sr{} are produced at higher latitudes, where the amount of ejecta is significantly smaller or even negligible. 
Finally, for the spiral wave wind ejecta (bottom-right panel), matter ejection happens predominantly at low latitudes, but in this case the production of heavy nuclei and \he{} is more limited to the region across the equator ($70^\circ \lesssim \theta \lesssim 110^\circ$), \h{} is negligible everywhere (since $f_H(\theta)$ is maximum when $f_{\rm total}(\theta)$ is minimum), while light $r$-process elements, like \sr{}, trace well the overall ejecta distribution, with a significant excess/deficiency at very high/low latitudes, respectively.

\subsection{Nucleosynthesis: analysis}
\label{subsec:nucleosynthesis analysis}

\begin{figure*}[!t]
\centering
\includegraphics[width = \linewidth]{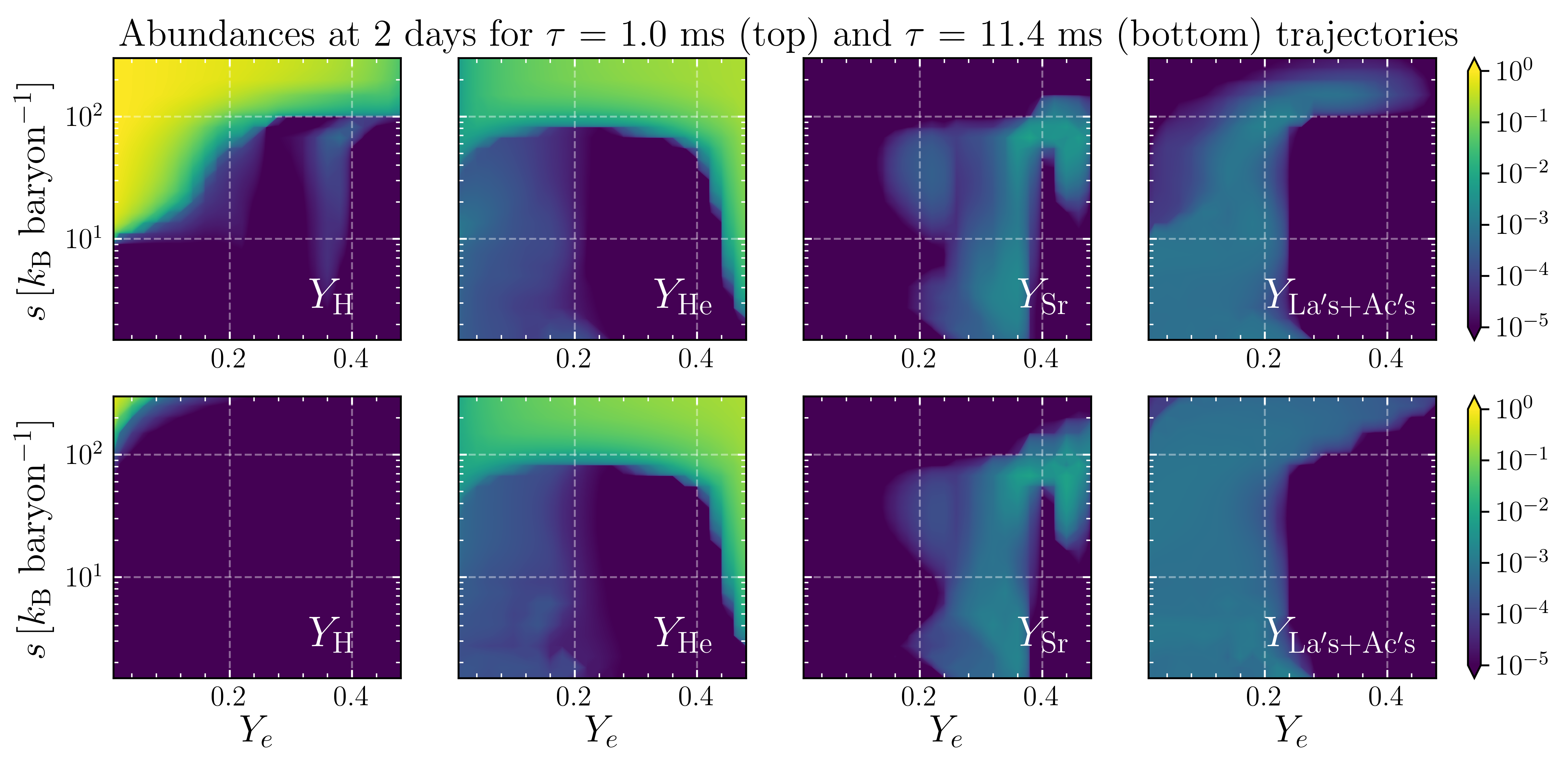} 
\caption{Number abundances of hydrogen, helium, strontium, and lanthanides and actinides summed together, as a function of $Y_e$ and $s$, for $\tau \sim 1.0~{\rm ms}$ (top) and $\tau \sim 10~{\rm ms}$ (bottom), at 2 days after merger, as obtained by our parametric nucleosynthesis calculations. }
\label{fig:abundances slices}
\end{figure*}

We investigate the origin of very light elements and of \sr{} first by presenting their abundances in \reffig{fig:abundances slices}, alongside with the sum of the lanthanides and actinides ones. These abundances were obtained by considering individual trajectories characterized by broad ranges of $Y_e$ and $s$, and two expansion timescales that bracket the relevant scales for dynamical and spiral-wave wind ejecta, namely $\tau = 1~{\rm ms}$ and $\tau = 11.4~{\rm ms}$, corresponding to $v_{\infty} \lesssim c $ and $v_{\infty}  \approx 0.1c$, respectively. Despite the fact that in \reffig{fig:abundances from NR simulations} we presented also the sum of the elements with $3 \leq Z \leq 19$, we do not report here their detailed abundances since they are subdominant and they do not produce any relevant feature on the presented scales.

\subsubsection{Hydrogen}
The presence of \h{} in the ejecta is related to high-$s$ and low-$Y_e$ matter that expands very rapidly, as visible in the top left panel of \reffig{fig:abundances slices} \citep[see also][]{Metz14,Ishii.etal:2018}.
In these conditions, \h~is produced as a decay product of free neutrons 
within a few tens of minutes. Indeed the ejecta at NSE freeze-out is composed mostly of free $n$'s, several percents in mass of $\alpha$ particles and very few heavier seed nuclei ($A \lesssim 100$). Due to the extremely fast density drop, a large fraction of $n$'s do not participate in the \rprocess~\citep{lipp15}. The larger the abundance of \h~is, the smaller the abundance of heavy elements is, with a narrow intermediate regime of comparable abundances. For $\tau \gtrsim 10~{\rm ms}$, \h~production becomes always negligible, unless for combinations of extremely high and low values of entropy and electron fraction, respectively.

\subsubsection{Helium}

\begin{figure*}
\centering
\includegraphics[width = \textwidth]{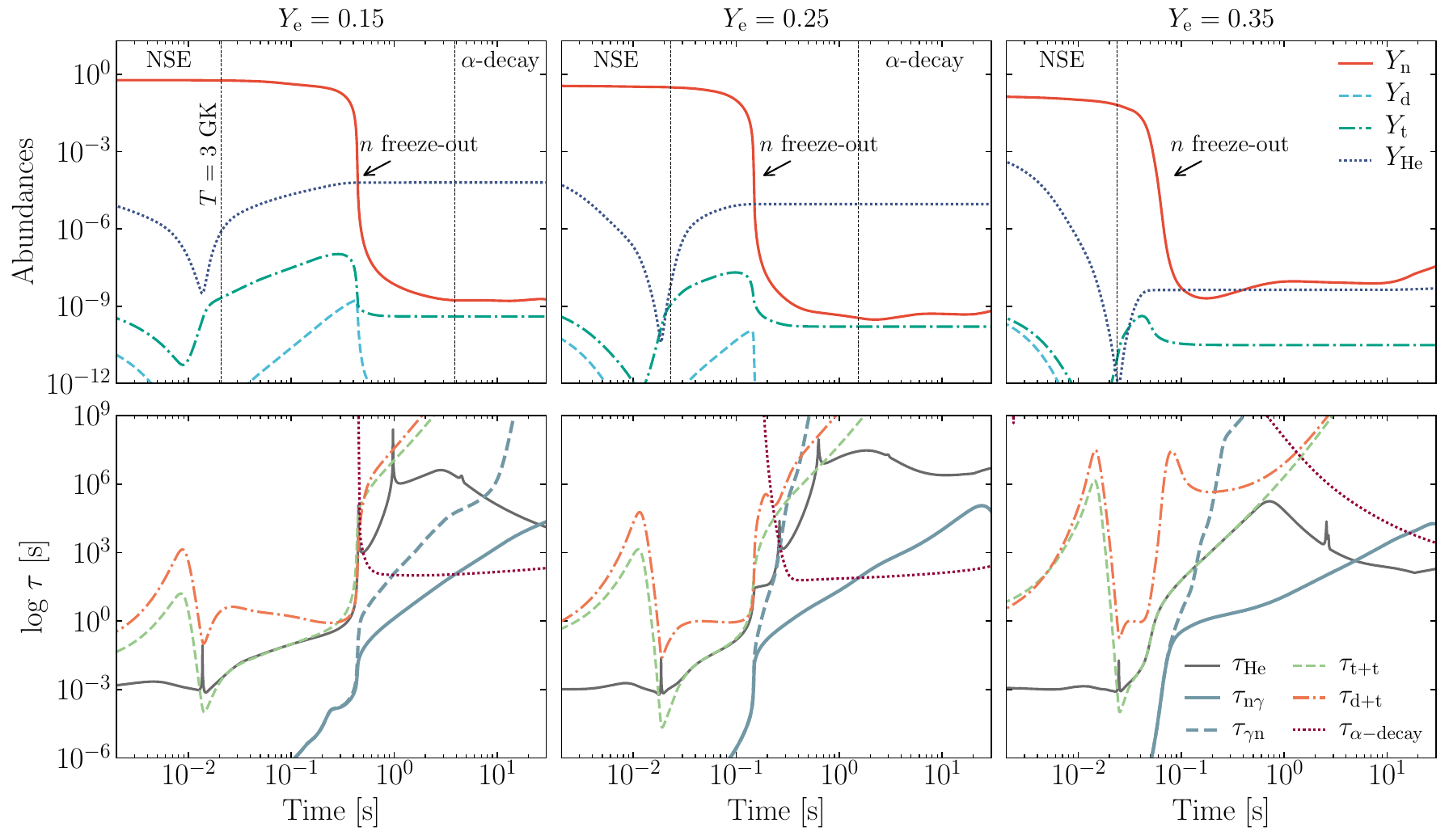} 
\caption{Evolution of a few selected abundances (top) and some relevant reaction timescales (bottom) for three low entropy trajectories. For all trajectories, $s=10~{k_{\rm B}~{\rm baryon^{-1}}}$ and $\tau=10~{\rm ms}$, while $Y_e=0.15,0.25,0.35$ moving from left to right. Before $n$ freeze-out, reactions involving deuterium and tritium account for $\alpha$ particle production, while after $\alpha$-decay of translead nuclei. 
Vertical dotted lines correspond to the time at which the temperature drops below 3 GK, roughly corresponding to the NSE freeze-out, and to the time when $\alpha$-decays become effective. The point when $Y_n = 10^{-4}$ is also presented as $n$ freeze-out.
For $0.2 \lesssim Y_e \lesssim 0.45$ the faster freeze-out and the lack of translead nuclei produce a lower $Y_{\rm He}$.}
\label{fig:tau}
\end{figure*}

As visible in \reffig{fig:abundances slices}, the production of \he{} can happen both in the absence of and in association with heavy elements, as also suggested by the simulation yields. The former case is realized for $Y_e \gtrsim 0.45 $, when most neutrons are locked inside strongly bound nuclei (including $\alpha$'s) at NSE freeze out, and the ratio between free neutrons and seed nuclei is not large enough to guarantee \rprocess~nucleosynthesis beyond the second peak. 
The latter case takes place in two different regimes:
i) at high entropy ($s \gtrsim 60 k_B~{\rm baryon^{-1}}$) for a broad range of electron fractions ($Y_e \lesssim 0.4$);
ii) at low entropy ($s \lesssim 60 k_B~{\rm baryon^{-1}}$) for relatively low electron fractions ($Y_e \lesssim 0.23$).
The high-$s$ regime was considered for many years as the main scenario for \rprocess~nucleosynthesis in supernova winds \citep[e.g.][]{Woosely1994,Farouqi2010,Arcones.Pinedo:2011}.
High-$s$ conditions favor the production of $\alpha$ particles and fewer heavier seed nuclei at NSE freeze-out  ($\alpha$-rich freeze-out).
The free neutrons are captured by the few seed nuclei to produce the heaviest elements, while for $Y_e \gtrsim 0.40$ the neutron-to-seed ratio becomes too small for this process to occur.
In the low-$Y_e$, low-$s$ regime, \he~production correlates with the production of heavy \rprocess~elements, particularly of actinides.
In these conditions, matter stays sufficiently dense while cooling to produce neutron rich iron group nuclei in NSE conditions. However, not all free neutrons are bound in nuclei so that they can be captured to produce heavy nuclei through the $r$-process.
To identify which processes are responsible for this correlation we focus on three representative low-$s$, low-$Y_e$ trajectories.
We fix, in particular, $s=10~k_B~{\rm baryon^{-1}}$ and $\tau=10~{\rm ms}$, while we consider $Y_{\rm e} = 0.15,0.25,0.35$. In the top panels of \reffig{fig:tau} we represent the abundances of a few selected isotopes as a function of time (measured with respect to the time when $T=5~{\rm GK}$), including free neutrons (n), deuterium (d), tritium (t), and $\alpha$ particles. Other \he{} isotopes are always subdominant by several orders of magnitude, and will be neglected in the following analysis. In the bottom panels we report some relevant timescales.
The lifetime of \he{} is computed as $
\tau_{\rm He} = \left| \left(  {\rm d} Y_{\rm He}/ {\rm d} t \right)/Y_{\rm He} \right|^{-1}
$.
Similarly, we introduce the average radiative neutron capture timescale per nucleus as:
\begin{equation}
   \tau_{\rm (n, \gamma)}=\frac{\sum\limits_{A, Z} Y_{(A, Z)}}{\sum\limits_{A, Z} Y_{(A, Z)} Y_{\rm n}\langle\sigma v\rangle_{(A, Z)}} \, ,
\end{equation} 
the average photodissociation timescale per nucleus as:
\begin{equation}
   \tau_{\rm (\gamma, n)}=\frac{\sum\limits_{A, Z} Y_{(A, Z)}}{\sum\limits_{A, Z} Y_{(A, Z)} \lambda_{{\rm \gamma},(A, Z)}} \, ,
\end{equation}
the $\alpha$ production timescale through ${\rm t + t \rightarrow n + n} + \alpha$  reaction as:
\begin{equation}
   \tau_{\rm t + t}=\frac{Y_{\alpha}}{Y_{\rm t}^2 \langle\sigma v\rangle_{\rm t + t}} \, ,
\end{equation}
the \he{} production timescale through ${\rm d + t \rightarrow n} + \alpha$ reaction as:
\begin{equation}
   \tau_{\rm d + t}=\frac{Y_{\alpha}}{Y_{\rm d}Y_{\rm t} \langle\sigma v\rangle_{\rm d + t}} \, , 
\end{equation}
 and finally the average $\alpha$-decay timescale per nucleus as:
\begin{equation}
   \tau_{\rm \alpha-decay}=\frac{\sum\limits_{A, Z} Y_{(A, Z)}}{\sum\limits_{A, Z} Y_{(A, Z)} \lambda_{{\rm \alpha},(A, Z)}} \, .
\end{equation}
In the previous expressions $Y_{(A, Z)}$ is the number abundance of a nucleus with atomic number $Z$ and mass number $A$, $\langle\sigma v\rangle_{Z, A}$, $\lambda_{{\rm \gamma},(A, Z)}$, and $\lambda_{{\rm \alpha},(A, Z)}$ are the corresponding neutron capture rate, photodisintegration ${\rm (\gamma, n)}$ rate, and $\alpha$-decay rate, respectively, $\langle\sigma v\rangle_{\rm t + t}$ represents the ${\rm t+ t}$ reaction rate, while $\langle\sigma v\rangle_{\rm d + t}$ the ${\rm d + t}$ reaction rate.
Helium is initially depleted to form heavier iron group nuclei in expanding matter under NSE conditions, up to the point when charged-particle freeze-out occurs ($T \gtrsim 3$ GK). Despite the initially larger $\alpha$ abundance, in the $Y_e=0.35$ trajectory, due to the lower neutron abundance, (n, $\alpha$) reactions are inefficient in producing \he{} that is instead effectively destroyed through ($\alpha$, n) reactions, thus reducing $Y_{\rm He}$ more significantly than for smaller $Y_e$'s.
After NSE freeze-out, ${\rm (n,\gamma)}$-${\rm (\gamma,n)}$ equilibrium guarantees a high neutron density and neutron-to-seed ratio, and starting from seed nuclei with $A \lesssim 100$ drives the formation of heavier nuclei through the $r$-process nucleosynthesis, far from the valley of stability. The high abundance of free neutrons provides also an almost steady supply of free protons (through $n$-decay) and thus the efficient formation of d and t. Reactions such as ${\rm t + t \rightarrow n + n +{}^4 He}$ and ${\rm d + t \rightarrow n +{}^4 He}$ are not in equilibrium with their inverse. As a consequence \he{} nuclei accumulate and increase their abundance till ${\rm (n,\gamma)}$-${\rm (\gamma,n)}$ equilibrium freezes-out (visible in the figure when $Y_n \lesssim 10^{-4}$). When $Y_{\rm n}$ drops, tritium and deuterium are no more produced 
and the production of \he~is halted. As visible in the lower panels of \reffig{fig:tau}, $\tau_{\rm He}$ is accounted by $\tau_{\rm d + t}$ and especially $\tau_{\rm t +t}$ in the time window between the end of NSE and the drop of $Y_{\rm n}$, clearly demonstrating that these two reactions are the main \he{} production channel.
At later times ($t \gtrsim$ 2 s), $\alpha$-decay of translead nuclei (if produced) becomes significant and $Y_{\rm He}$ increases further.
Lower initial $Y_e$ (i.e., a larger $Y_{n} \approx 1-Y_e$) results in:
i) a wider time window over which t and d can be efficiently produced and converted into \he{}; ii) a larger abundance of $\alpha$-decaying translead nuclei. 
This picture is further confirmed by the fact that if  $n$ decay is artificially removed from \texttt{Skynet}
$Y_{\rm He}$ stays initially frozen, before increasing only at $t \gtrsim$ 2 s due to $\alpha$ decays. When present ($Y_e \lesssim 0.2$), $\alpha$ 
decays can change \he{} abundance by a factor of 2 on a timescale of several days after merger, comparable to the kilonova timescale.
The relative importance between charged reactions and $\alpha$-decays has been analyzed in more detail in \refapp{appendix: alpha}.

\subsubsection{Elements from Lithium to Potassium}
For all elements between Lithium and Potassium, the predicted abundances are usually below $10^{-5}$ over the whole parameter space. Only for a few elements (namely, Beryllium, Nitrogen, Oxygen and Neon), abundances can be slightly above $10^{-5}$ in corners of the parameter space (usually, $Y_e \lesssim 0.3$ and high entropy). In the case of high entropy ejecta, at NSE freeze-out the densities are rather low ($\gtrsim 10^5 {\rm g~cm^{-3}}$). In the resulting $\alpha$-rich freeze-out conditions, the building of light elements through $\alpha + \alpha + n$ and triple-$\alpha$ reactions is very inefficient (unless the density of free neutrons can partially compensate). In the case of low entropy and possibly low $Y_e$ conditions, the nuclei distribution at NSE freeze-out is dominated by neutron-rich seed nuclei around the iron group (with $A \sim 70-90$). This distribution extends also towards lower mass numbers, but it is lower-bounded by the $N=20$ and $Z=20$ magic numbers. The subsequent $r$-process nucleosynthesis produces nuclei with $A \gtrsim 40$. For low electron fractions, $N$ can be significantly larger than $Z$, but the high neutron-to-seed ratio drives the nucleosynthesis far from calcium isotopes. For relatively large $Y_e$ ($Y_e \gtrsim 0.4$), the starting seeds are close to the valley of stability and $\beta$-decays are not effective in significantly lowering the atomic number. All this prevents the efficient formation of nuclei below calcium and other than \h{} and \he{} both in high and low entropy $r$-process nucleosynthesis.

\subsubsection{Strontium}
\reffig{fig:abundances slices} shows how the production of elements of the first $r$-process peak and immediately above it (as \sr) crucially depends on the ejecta $Y_e$. Indeed, for $0.2 \lesssim Y_e \lesssim 0.38$ \sr{} is robustly produced for entropy lower than $\sim 100~{k_{\rm B}~{\rm baryon^{-1}}}$. In these conditions, a weak \rprocess~nucleosynthesis occurs: at NSE freeze-out, 
typical seed nuclei are in the mass range of $A = 50-80$
and the neutron-to-seed ratio is of a few tens.
The subsequent neutron captures produce nuclei up to $A \sim 90$,
including \sr{}, but the nucleosynthesis does not reach the second \rprocess~peak. 
Within this regime, a significant increase of $Y_{\rm Sr}$ is observed only for $0.35 \lesssim Y_e \lesssim 0.38$, while at relatively large entropy ($20 \lesssim s [k_{\rm B}~{\rm baryon^{-1}}] \lesssim 80$) \sr{} is produced with a lower threshold $Y_e$ ($\lesssim 0.20$).
For $s \gtrsim 100 {k_{\rm B}~{\rm baryon^{-1}}}$ or $Y_e \gtrsim 0.48$, $\alpha$-rich freeze-out occurs. \sr{} can be also produced in $\alpha$-rich freeze-out conditions \citep[see, e.g.,][]{hoff97,Freiburghaus:1999}, but only for $20 \lesssim s [k_{\rm B}~{\rm baryon^{-1}}] \lesssim 120$ and $Y_e \gtrsim 0.4$. Indeed, for lower entropy the neutron-to-seed ratio decreases too much and the nucleosynthesis stops at the first \rprocess{} peak, while for larger entropy the density at NSE freeze-out is too low to allow 3 body reactions and most of the $\alpha$ particles do not further interact.

\subsection{Ejecta conditions}
\label{subsec:results ejecta}
\begin{figure*}[ht!]
\centering
\includegraphics[width = \textwidth]{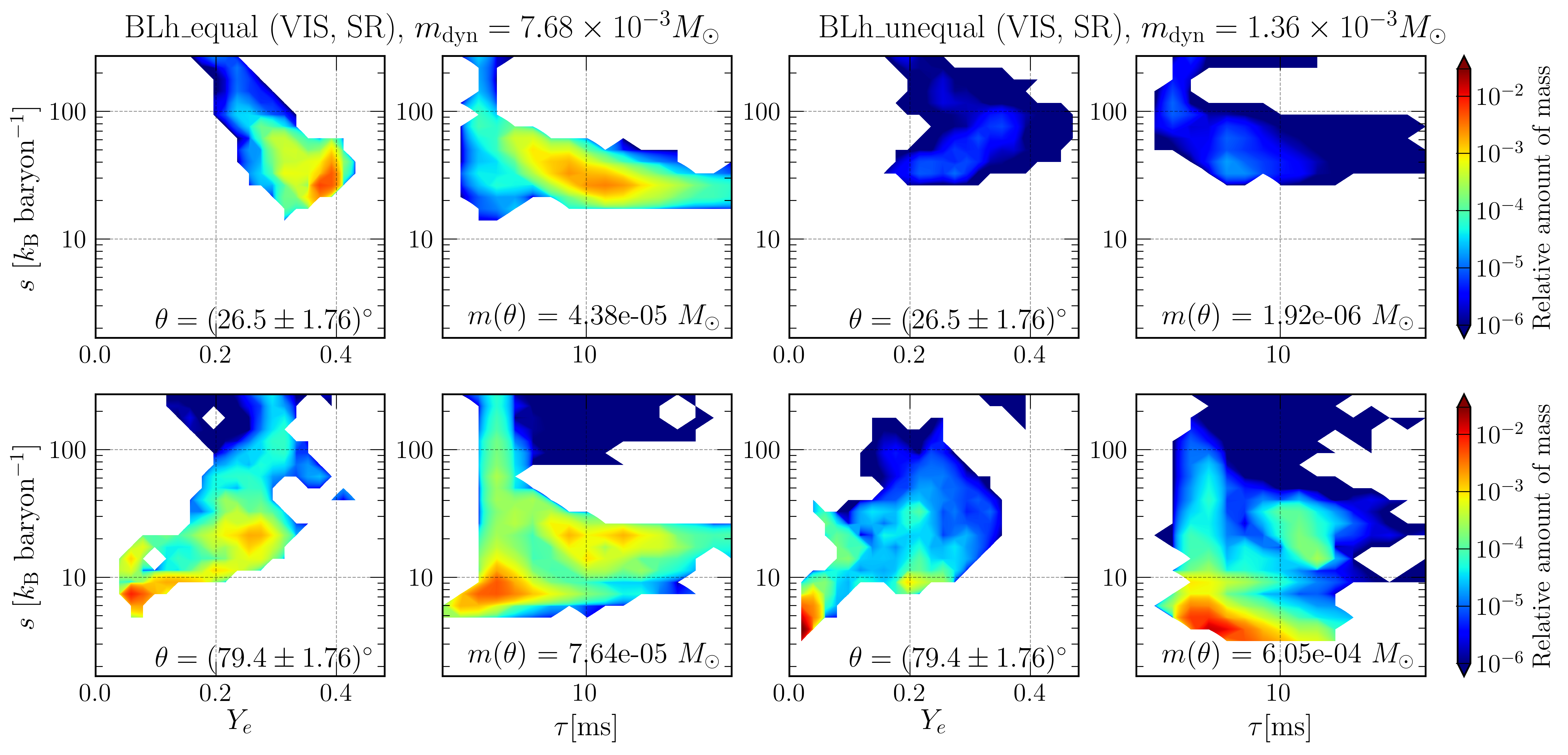}
\caption{Mass weighted, normalized histograms of the $(Y_e,s,\tau)$ conditions for the dynamical ejecta extracted from the SR, viscous simulation of the BLh\_equal model (left four panels), and of the  BLh\_unequal model (right four panels). For each simulation, we distinguish between the $(Y_e,s)$ (left panel) and $(\tau,s)$ (right panel) planes, and a high (top panel) and a low (bottom panel) latitude angle.}
\label{fig:histograms}
\end{figure*}
\begin{figure*}[ht!]
\centering
\includegraphics[width = \textwidth]{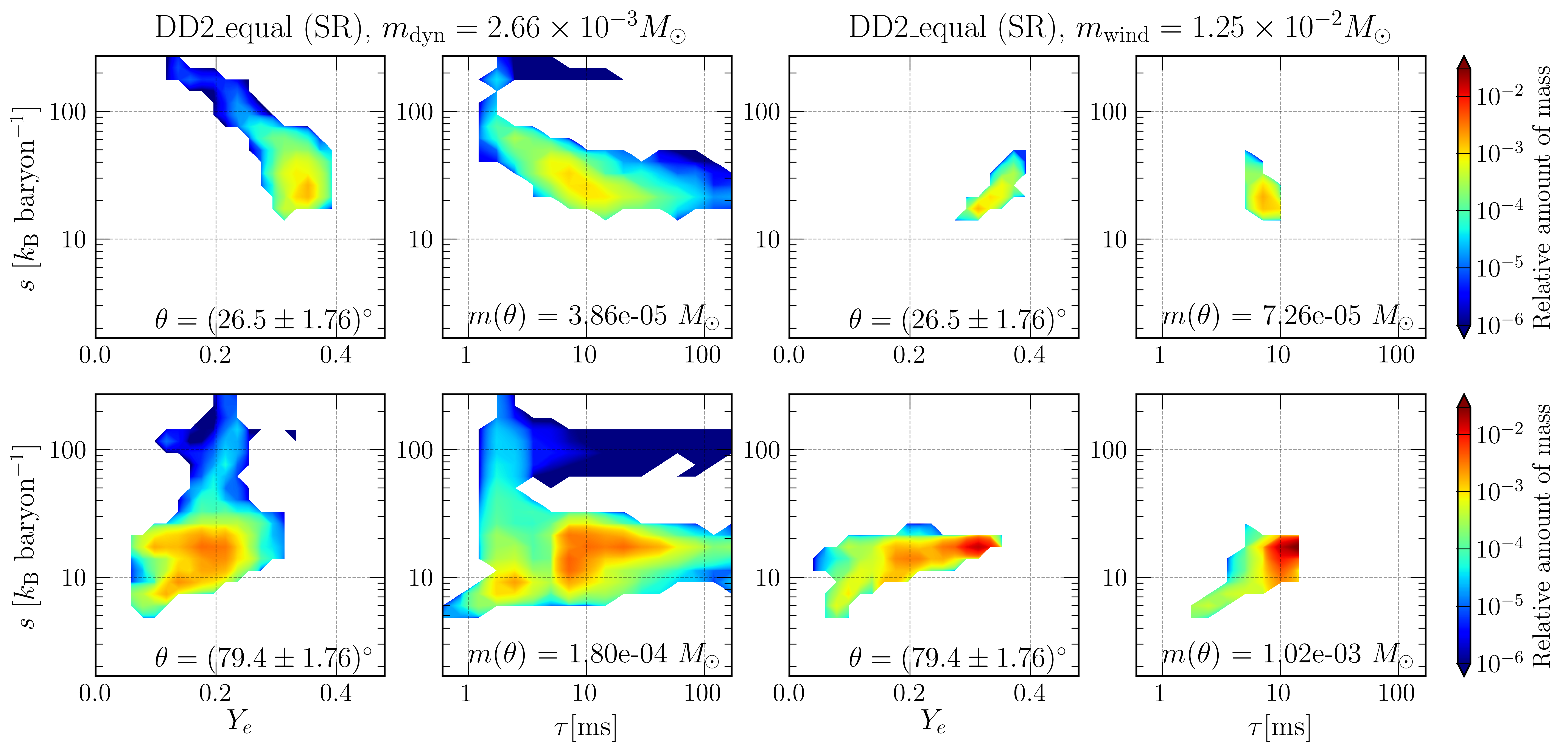}
\caption{Same as in \reffig{fig:histograms}, but for a SR, non-viscous simulation of the DD2\_equal model, employing the stiffer HS(DD2) EOS. In this case, we consider both dynamical ejecta (left four panels) and spiral wind ejecta (right four panels).}
\label{fig:histograms2}
\end{figure*}

We are now in the position to understand in which dynamics and thermodynamics conditions the abundances presented in \reffig{fig:abundances from NR simulations}  
have been synthesized by convolving the thermodynamic distribution of the ejecta from the simulations with the distribution of abundances computed as a function of the ejecta conditions, as visible in \reffig{fig:abundances slices}.
In \reffig{fig:histograms} and \reffig{fig:histograms2} we present the normalized, mass weighted histograms as extracted from three of the simulations considered in this work, one for each BNS model. We tested that the overall features discussed below depend neither on the resolution nor on the inclusion of physical viscosity. For each angular bin, the 3D distributions have been marginalized either with respect to $\tau$ (first and third columns) or $Y_e$ (second and forth columns). Since the ejection of matter is usually anisotropic, we consider two representative polar angles, one close to the polar axis (top panels) and one to the equatorial plane (bottom panels), while we integrate along the azimuthal direction.
The bulk of the ejecta have low entropy ($s < 40~k_{\rm B}~{\rm baryon^{-1}}$) and is very neutron rich, with equatorial ejecta being usually characterised by lower $Y_e$ and lower entropy. However, at both angles a high entropy tail (with $s > 60~k_{\rm B}~{\rm baryon^{-1}}$) expanding at high speeds ($\tau \lesssim 5 {\rm ms}$) is visible. This high-entropy, high-velocity tail in the ejecta is a signature of the so-called shock heated ejecta, produced by the bouncing remnant. This tail is more important for softer nuclear EOS, producing more violent mergers and stronger shocks.
In the case of very different colliding NS masses the properties of the ejecta are qualitatively different. In this case, the dynamical ejecta are mainly produced by the tidal disruption of the lightest NS.
Polar ejecta are almost absent and the equatorial ejecta are dominated by low entropy, low $Y_e$ matter that expand with $\tau \sim 10{\rm ms}$. The high-entropy, high-velocity tail in their ejecta distribution is almost absent.

For equal mass mergers, in the case of the softer BLh EOS most of the \h~and \he{} are efficiently synthesized in the high-$s$, fast expanding tail of the shock-heated component of the dynamical ejecta, while the subdominant \he{} synthesized in low-$s$, low-$Y_e$ conditions roughly traces the less abundant heavy $r$-process element distributions.
For the stiffer DD2 EOS, the merger is less violent. The high-$s$ tail of the shock-heated dynamical ejecta is thus less relevant and its contribution to the \he~production becomes comparable to the low-$s$, low-$Y_e$ contribution. \h~is still produced, but slightly less efficiently.
For the same merger model, the spiral wind ejecta have a pronounced, narrow peak in the velocity-entropy space, around $\tau \approx 8~{\rm ms}$ and $s=20 ~{k_B}~{\rm baryon^{-1}}$, and a broad $Y_e$ distribution with a peak around $\gtrsim 0.3$, but extending down to $0.1$. Thus, the production of \h~is suppressed while \he~is synthetized less efficiently than in the dynamical ejecta and in association with the more abundant lanthanides and actinides.
For the unequal mass case, because of the lack of the high entropy tail, low-$s$, low-$Y_e$ matter is the main source of \he, tracing the presence of
heavy \rprocess~elements, more abundant than \he{} by a few.\\
Strontium is synthesized in the high latitude dynamical and in the spiral-wave wind ejecta. In these ejecta, neutrino irradiation plays a fundamental role in increasing $Y_e$ above the production threshold. The larger entropy and $Y_e$ obtained in the case of spiral-wave winds or softer EOS enhance its production by a factor of a few. 
On the contrary, the equatorial ejecta that characterize very unequal mass merger are efficiently shielded from neutrino irradiation, preventing \sr{} production.

\subsection{Kilonova spectra}
\label{subsec:results kilo spectra}

\begin{table*}[ht!]
    \centering
    \begin{tabular}{c|c|c|c|c|c|c|c}
         \hline\hline
         Time	& $m_{\rm H}$ or $m_{\rm He}$ & $f_{\rm tr}$ &	
         $L_{\rm ph}$ & $v_{\rm ph}$ &
         (N)LTE & Notes & Effects \\ {~}	
         $[{\rm days}]$ & $[10^{-6} M_{\odot}]$ & $[-]$ & $\left[ 10^{41}{\rm erg~s^{-1}}\right]$ & $[c]$ & {~} & {~} \\ \hline \hline     
         \multicolumn{8}{c}{Hydrogen}  \\ \hline
         0.2-0.3 & 1.0 & 1.0 & 16 & 0.3 & LTE & baseline & none \\
         0.2-0.3 & 1-20 & 1.0 & 16 - 320 & 0.3 & LTE & sensitivity & none \\
         \hline \hline
         0.2-0.3 & 1.0 & 1.0 & 16 & 0.3 & NLTE & baseline & none \\
         0.2-0.3 & 1-20 & 1.0 & 1-320 & 0.3 & NLTE & sensitivity & H$\alpha$ line for $L_{\rm ph} \lesssim 2.15$ and $m_H \gtrsim 10$ \\
         \hline \hline
         \multicolumn{8}{c}{Helium}  \\ \hline
         2 & 4 & 0.52 & 4 & 0.25 & LTE & baseline &  none \\
         2 & 4-80 & 0.51 & 4-80 & 0.25 & LTE & sensitivity & none \\ \hline
         3 & 4 & 1.0 & 4 & 0.235 & LTE & baseline & none \\
         3 & 4-80 & 1.0 & 4-80 & 0.235 & LTE & sensitivity & none \\ \hline \hline
         2 & 4 & 0.52 & 4 & 0.25 & NLTE & baseline & none \\
         2 & 4-80 & 0.51 & 0.2-80 & 0.25 & NLTE & sensitivity &  He I feature for $0.27 \lesssim L_{\rm ph} \lesssim 0.96$ and $m_{\rm He}=4$  \\ \hline
         3 & 4 & 1.0 & 4 & 0.235 & NLTE & baseline & none \\
         3 & 4-80 & 1.0 & 0.2-80 & 0.235 & NLTE & sensitivity & He I feature for $m_{\rm He} > 40$ and $L_{\rm ph}=4$ \\
         ~& ~& ~& ~& ~& ~ & ~ & He I feature for $0.48 \lesssim L_{\rm ph} \lesssim 1.5 $ and $m_{\rm He}=4$ \\
         \hline \hline
    \end{tabular}
    \caption{Summary of the kilonova spectra results obtained using \texttt{TARDIS}. At early times after merger (top, $0.2-0.3~{\rm days}$) we consider tiny kilonova atmospheres rich in \h{}, while at later times (bottom, $2-3~{\rm days}$) larger atmospheres with \he{}. For each time, we consider a baseline model characterized by a certain amount of \h{} or \he{}, $m_{\rm H}$ and $m_{\rm He}$, a certain transparency factor, $f_{\rm tr}$, corresponding to the fraction of the element mass outside the photosphere, a photospheric luminosity and velocity, $L_{\rm ph}$ and $v_{\rm ph}$, respectively. We study the sensitivity of our results with respect to LTE VS NLTE, and with respect to the element masses and photospheric luminosity.}
    \label{tab:summary spectra}
\end{table*}

To test whether the \h{} and \he{} synthesised in the early ejecta of BNS mergers can produce features, we compute synthetic kilonova spectra using \texttt{TARDIS}, initialized with the ejecta properties derived in our models.
We consider a fiducial ejecta model characterized by $1.5 \times 10^{-3} M_{\odot}$ of dynamical ejecta and $1.95 \times 10^{-2} M_{\odot}$ of spiral wave wind ejecta (obtained by considering $R_{\rm wind} = 0.15 M_{\odot} s^{-1}$ acting for $\Delta t_{\rm wind} = 0.13 {\rm s}$). 
The dynamical ejecta expand at an average speed of $0.22c$, while 
the spiral wave wind ejecta at $0.15c$.
We assume that the two ejecta have interacted producing a single homologously expanding profile, which however retains information about
the ejecta stratification in the composition (i.e. we assume no large scale mixing on the kilonova timescale). 
The expanding ejecta are thus described by the profile \refeq{eq: density homologous expansion}, where $M$ is the sum of the two ejecta masses and $v_{\rm avg}$ is obtained by imposing linear momentum conservation, i.e. it is the mass weighted speed. We assume $10^{-6} M_{\odot}$ of \h{}, located at the top head of the ejecta and moving at $\sim$ 0.33$c$. We note that this speed is $\sim 30\%$ lower than the speed of the fast expanding tail of the dynamical ejecta as extracted from our simulations at $\sim$10ms after merger. This is a consequence of the adoption of a single profile for the whole ejecta at the kilonova timescale. Since this produces a possibly denser \h{} layer, the following analysis should be intended as an upper limit on the explored effects. We further assume that \h{} is
produced in association with heavy elements with a mass fraction of $X_{\rm H} = 0.5$  \citep[see e.g.][]{Metz14}. Below it, we consider a mixture of $r$-process elements, inside which \he{} is uniformly distributed inside the top part of the ejecta, corresponding to the dynamical ejecta only (i.e., we consider no \he{} in the underlying spiral wave wind ejecta). For the sake of concreteness, we consider low latitude emission (to which a larger solid angle is associated) and we assume $\kappa=10~{\rm cm^2~g^{-1}}$. Based on \refeq{eq: photosphere} and \refeq{eq: transparent mass}, we estimate the amount of ejecta (and thus of \h{} and \he{}) in optically thin conditions and the speed of matter at the photosphere. Additionally, we consider representative values of the luminosity at the photosphere, based on time-dependent bolometric luminosity models fitted against AT2017gfo \citep{Smartt:2017fuw}, ranging from $1.6 \times 10^{42}{\rm erg~s^{-1}}$ a few hours after merger to $4 \times 10^{41}{\rm erg~s^{-1}}$ at 2-3 days. The set of all these values are considered as baseline values, while we test the robustness of our results both with respect to the \h{} and \he{} masses, and to the bolometric luminosity. The results of our investigation are reported in \reftab{tab:summary spectra}. 

We first consider the spectra obtained by considering LTE conditions in the ejecta. For all explored configurations, i.e. starting from the baseline models and increasing the \h{} or the \he{} mass, and the photosphere luminosities up to a factor of 20, we do not observe any observable feature in the spectrum at $\sim$ 5 and 8 hours, and at 2 and 3 days for \h{} and \he{}, respectively.
We choose these times because the range 0.2-0.3 days is the time interval when we expect the outer \h{}-rich shell to become optically thin. For helium, we investigate 2 and 3 days after merger, since these are the times when we expect $\sim 50\%$ and $\sim 100\%$
of \he{} to be above the photosphere in our models.
These results can be understood by considering that for the density and temperature conditions expected during a kilonova and assuming LTE, \h{} and \he{} recombine to atomic form very rapidly, see \refapp{appendix: saha}. Only for slowly expanding ejecta, ionized states are present up to a few hours after the merger.
In fact, while a few hours after explosions the ejecta temperature reaches $10^5$~K, already at 0.5 day the temperature drops to $10^4$~K. In these conditions, the small predicted masses of \h{} and \he{} are not expected to produce persistent lines. 
These results depend very weakly on the specific trajectory.

\begin{figure*}
\centering
\includegraphics[width=\linewidth]{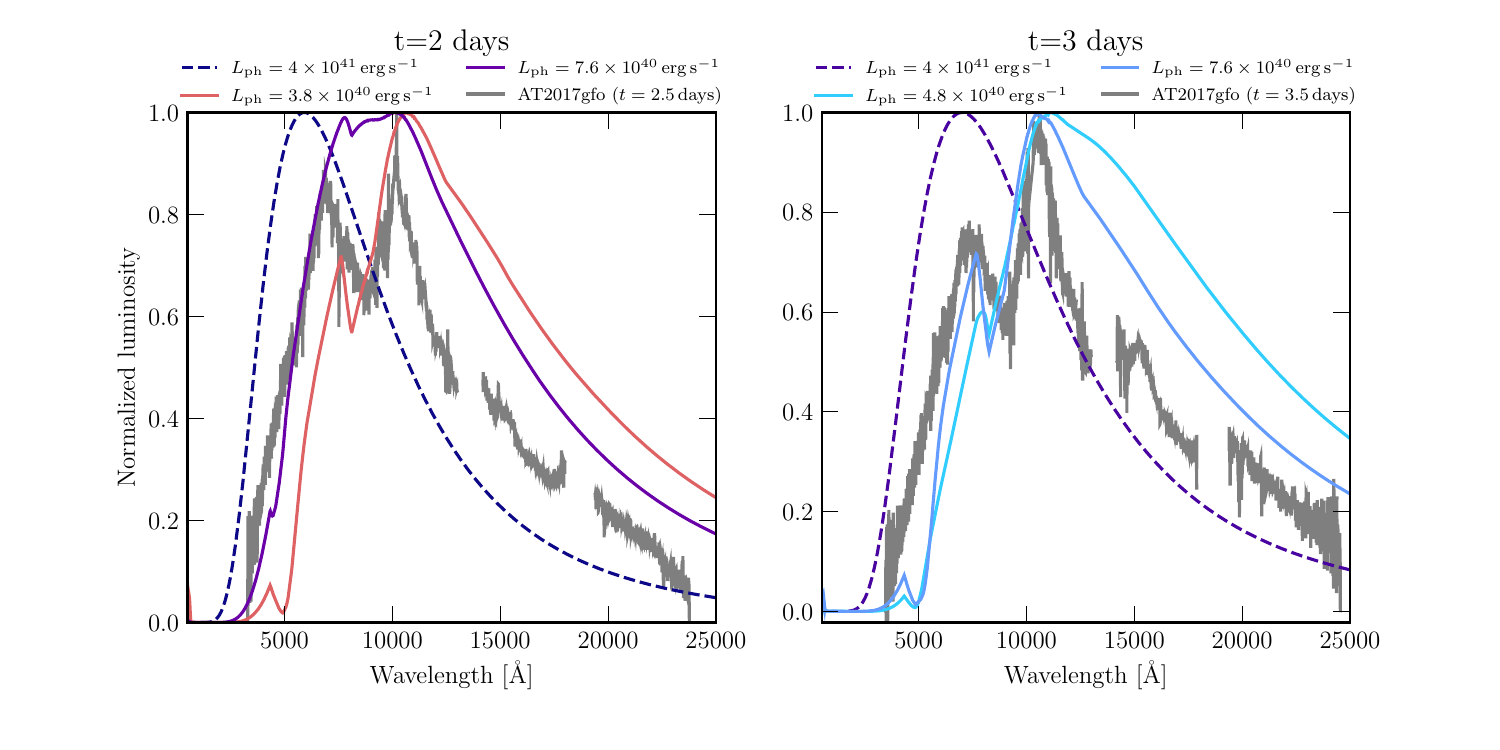}
\caption{Synthetic \texttt{TARDIS} spectra including NLTE effects for a kilonova observed along the equatorial direction, for different luminosities and with an atmosphere enriched in \he{}. The left (right) panel refers to the spectrum at 2 (3) days after merger. The dashed lines correspond to our baseline models, the solid lines to models for which the luminosity is such that NLTE effects appear as a He I $\lambda$ 10831 line. The gray lines correspond to the spectra of AT2017gfo at 2.5 days (left) and 3.5 days (right) after the removal of the telluric lines \citep{pian17}.}
\label{fig:tardis lum dep He NLTE}
\end{figure*}

We explore the possible impact of NLTE excitations by using the analytical approximation for the NLTE \he{} level population developed inside \texttt{TARDIS} for the \he{} rich ejecta of double detonation type Ia SNe \citep{boyle17}. The predicted \he{} line strength changes dramatically with the NLTE treatment. In these explorations, we also test the case of photospheric luminosities lower than the baseline value. While for the fiducial values of \he{} mass and luminosity no clear features can be observed,
a larger amount of He can produce a visible broad \he{} I $\lambda$ 10831 line. In particular, using our baseline photospheric luminosity at two days ($4 \times 10^{41}~{\rm erg~s^{-1}}$), an absorption feature becomes visible for $m_{\rm He} > 8 \times10^{-5} M_{\odot} $.
Lower masses (of the order of $4\times10^{-6}\,{M_\odot}$, comparable with our baseline value) of \he{} can produce the \he{} I $\lambda$ 10831 feature for days if the kilonova is fainter than our reference case. For example, at two days we obtained the same \he{} I $\lambda$ 10831 feature if the luminosity is between $2.7 \times 10^{40}~{\rm erg\,s^{-1}}$ and $9.6 \times 10^{40} {\rm erg\,s^{-1}}$, while at three days if   $L=3.8\times10^{40}-1.5\times10^{41} {\rm erg\,s^{-1}}$, see \reffig{fig:tardis lum dep He NLTE}.
These ranges in the photospheric luminosity can be understood since 
for too high luminosity \he{} is fully ionized and lines are not produced. 

We also checked the possible impact of NLTE effects for \h{}. 
None of the tested baseline configurations including NLTE effects in \texttt{TARDIS} as in \citet{vogl19} produce noticeable \h{} lines at optical wavelengths. For the baseline luminosity, at 0.3 days $10^{-5} M_{\odot}$ of \h{} (thus, ten times more than predicted by our models) are necessary to produce weak Lyman lines below 1000\AA. Balmer lines are visible only for an even (and probably implausible) larger amount of \h{}, $ \gtrsim 10^{-3} M_{\odot}$, with first hints at $ \sim 10^{-4} M_{\odot}$.
The temperatures are initially so high that there is very little neutral hydrogen. A smaller luminosity mitigates this effect, but the \h{} mass required to produce visible Balmer lines is still around  $10^{-5} M_{\odot}$.

\section{Discussion}\label{sec:discussion}

\citet{watson19} estimated between 1 and 5 $\times 10^{-5} M_{\odot}$ of \sr{} to explain the observed spectral feature attributable to the \sr{} II line in the early spectrum of AT2017gfo. As discussed in great detail in that work, this estimate must be taken with care due to a few simplifications contained inside the underlying kilonova model (for example, the usage of a spherically symmetric model, or of a single photosphere or the lack of possible NLTE effects). However, keeping in mind these important caveats, we can compare our finding with their inferred mass. On the one hand, the amount of \sr{} produced in the dynamical ejecta of our equal mass models is compatible with the inferred one, even when anisotropic production is taken into account and all \sr{} is assumed to be transparent within a few days. Indeed, since AT2017gfo was 
most probably observed from high latitudes ($15^\circ \lesssim \theta \lesssim 30^\circ$) and \sr{} is mainly produced at those latitudes in the dynamical ejecta, an equivalent spherically symmetric model would require about twice as much total dynamical ejecta ($f_{\rm Sr} \sim 2$ for $\theta \sim 20^{\circ}$, as visible for example in the top-left panel of \reffig{fig:abundances angular distribution}).
On the other hand, in the case of very unequal mass models, \sr{} is produced in a more isotropic way. The required \sr{} mass is however significantly larger than the one obtained in the dynamical ejecta of our corresponding simulations.
In addition to the dynamical ejecta, for models that do not form quickly a BH, the spiral wave wind is an efficient mechanism to unbind matter from the disk \citep{ned19,ned20}. For example, assuming an emission rate of $0.15 M_{\odot}~{\rm s^{-1}}$, a timescale of $\Delta t_{\rm wind} \sim 0.13 {\rm s}$ would be required to unbind $\sim 0.02 M_{\odot}$ \citep[this is the amount usually required by kilonova models to explain the blue component of AT2017gfo, see e.g.][]{kas17,Cowperthwaite.etal:2017,Villar.etal:2017,Perego:2017wtu,Rosswog:2017sdn,Breschi.etal:2021}. Based on the total \sr{} mass fraction obtained in our models, these ejecta would translate to $4-8 \times 10^{-4} M_{\odot}$ of additional \sr{}. However, the production of \sr{} is very asymmetric in this wind and, in particular, for a high latitude viewing angle ($\theta \sim 20^{\circ}$), the amount of \sr{} of an equivalent spherically symmetric model should be about 20\% of the its actual amount, meaning $8-16 \times 10^{-5} M_{\odot}$ of additional \sr{} in the spiral wave wind ejecta, as visible in the bottom-right panel of \reffig{fig:abundances angular distribution}. This amount is still larger than the inferred one, but only by a factor of a few. While the many uncertainties in the spectrum calculation and the assumption that all \sr{} was above the relevant photosphere at the time of interest for the spectrum can possibly weaken this discrepancy, our results may suggest that the timescale over which a spiral wave wind could have been active in GW170817 remnant should be not significantly larger than 100ms, otherwise the corresponding ejecta would imply a much larger amount of \sr{} than the one reported by \citet{watson19}.

In \reffig{fig:tardis lum dep He NLTE}, we also report the 2.5 and 3.5 days spectra of AT2017gfo \citep{pian17} showing a broad absorption at 810 nm that was explained by a blue-shifted transition of \sr{~II} \citep{watson19}.
Interestingly, based on the velocity profile only, this feature could be consistent also with the \he{} 10831 line in the ejecta expanding at $0.25c$. However, the lower luminosities (with respect to 
the one deduced by AT2017gfo spectra assuming a distance of $d_\mathrm{L}=40\,\mathrm{Mpc}$) required to observe significant NLTE effects seem to disfavor the interpretation of this feature as caused by \he{} rather than \sr{}. More luminous kilonovae, more compatible with AT2017gfo and closer to our baseline model, require \he{} masses significantly larger than our baseline value ($\gtrsim 10$) to produce similar effects. This discrepancy is possibly mitigated by the anisotropy in the ejection of $\he{}$ at polar latitudes, see \reffig{fig:abundances angular distribution}, but only by a factor of $\sim 2$, if we consider $\theta \sim 20^{\circ}$.

In addition to producing \h{}, the decay of free neutrons at the forefront of the dynamical ejecta can power an electromagnetic precursor of the kilonova \citep{Metz14}. However, the amount of free neutrons available to decay and to power such a precursor (which can be identified with the amount of \h{} in the dynamical ejecta, see \reftab{tab:summary dynamical}) is between one and two orders of magnitudes smaller than the one used by \citet{Metz14} and obtained by \citet{just15}. We stress that neutrino effects were not considered in those simulations, while their impact on the amount of free neutrons is crucial, as visible in our results and as also discussed by \citet{Ishii.etal:2018} and \citet{Manu.etal:2020}. 
Indeed, our results agree within a factor of a few with those reported in these latter papers. Since the precursor luminosity is expected to be proportional to the free neutron mass, the predicted magnitude in the UV bands is likely to be more than 3.5 magnitudes smaller than the one suggested in \citet{Metz14} even within the first hour after merger \citep[where the peak is expected to occur, as suggested by][]{Ishii.etal:2018} . For those conditions, it is unclear if this contribution will be visible or if it will be dominated by the rising kilonova emission.

Dynamical and spiral-wave wind ejecta from BNS mergers are not the only environment where \h{} and \he{} can be synthesized in association with \rprocess~elements.
In the case of BH-NS mergers, dynamical ejecta conditions are very similar to the ones observed in very unequal BNS mergers, with possibly larger average expansion velocities \citep[e.g.][]{just15,rob17,fer17,kyu18}. Thus, in this case we expect a significant \he{} production in association with heavy \rprocess~nucleosynthesis, as visible in some of the nucleosynthesis results of the above works.
Since \h{} is significantly produced in the high speed tail of the dynamical ejecta and these ejecta are mainly produced in shock heated conditions, \h{} is expected to be even less abundant in BH-NS mergers. 
Both in the case of BNS and BH-NS mergers, neutrinos, viscosity and magnetic processes can drive matter ejection on the viscous timescale expanding at significantly smaller velocity than the dynamical ejecta, $v_{\infty}\lesssim 0.1c$ \citep[see e.g.][]{metz09,fer13,per14,Metzger:2014ila,sie14,Martin:2015hxa,just15,lip17b,Wu.etal:2016,rad18b,fuj18,fer19,Miller:2019dpt}.
The possible presence of \he{} in these ejecta and their potential spectroscopical relevance was already anticipated by \cite{fer13}, based on analytical estimates \citep{hoff97}.
Detailed numerical simulations \citep[see e.g.][]{fer13,just15,fuj18} show
the possible presence of a significant high-entropy tail in the ejecta distribution, especially close to the polar axis, where the production of \he{} in $\alpha$-rich freeze-out conditions occurs. This tail is particular prominent in the case of GRMHD simulations \citep{fer19}. 
For BNS mergers, if the central massive NS survives on a timescale comparable to or larger than the viscous timescale, $\nu$-irradiation can increase $Y_e$ even above 0.45, producing efficiently \he{} in $\alpha$-rich freeze-out conditions. Thus, also these ejecta can host significant \he{} production \citep{lip17b}. We stress, however, that these slower ejecta is expected to become transparent at later times, when the presence of heavy elements and stratified ejecta can substantially increase the spectrum complexity.

\section{Conclusions}\label{sec:conclu}

In this paper, we have analyzed the production of very light elements (between \h{} and K) in the early ejecta expelled by binary neutron star mergers, and we have investigated their detectability in kilonova spectra. In particular we have focused on the nucleosynthesis occurring in the dynamical and spiral wave wind ejecta obtained by detailed numerical relativity simulations targeted to the GW170817 event, and we have explored their dependence on the mass ratio and nuclear EOS.

We found that \h{} and \he{} can be robustly synthesized in the dynamical ejecta, with a mass fraction ranging between $10^{-3}$ and $10^{-2}$, while their production is negligible in spiral wave wind ejecta.
The total amount of \h{} ranges between $\sim 0.5$ and $2 \times 10^{-6} M_{\odot}$, while the one of \he{} between $\sim 2$ and $10 \times 10^{-6} M_{\odot}$.
Hydrogen is mostly produced in the high speed tail of the dynamical ejecta  \citep[see e.g.][]{just15,Ishii.etal:2018} while the production of \he{} can happen both in high-entropy, high-$Y_e$ and in low-entropy, low-$Y_e$ conditions. In the latter case, the presence of \he{} is associated with the production of heavy \rprocess~elements and its amount can increase by a factor of $\sim$2 during the kilonova timescale, due to $\alpha$ decays. With the exception of the \he{} produced by $\alpha$ decay of very heavy elements, neutron decay is the driving nuclear process behind the nucleosynthesis of both these elements. Indeed, in addition to producing \h{} from unburned free neutrons on timescales comparable to the free neutron lifetime, the early (i.e. within the first second) decay of free neutrons in very neutron rich environments produces deuterium and tritium that immediately fuse to produce \he{}.

Based on the masses and on the ejecta properties computed in our model, we produced synthetic spectra with \texttt{TARDIS} to test whether dynamical ejecta enriched in \h{} or \he{} can leave an observational signature in the observed kilonova spectrum. Our analysis suggests that no effects are visible if LTE conditions are assumed. NLTE effects are required to produce a significant He I $\lambda$ 10831 spectral feature at $t\gtrsim2$ days after merger, while they do not produce any clear H feature for our fiducial configurations.
Then, we further explored the luminosity and mass parameter spaces to define admissible ranges allowing for the He I $\lambda$ 10831 spectral feature in the spectrum. Considering the baseline luminosity, a \he{} mass at least one order of magnitude larger than our fiducial value is required to produce a significant feature.
However, a lower \he{} mass, comparable to our reference value, could be sufficient to produce an observable feature between 2 and 3 days if the kilonova is fainter than our reference value by one order of magnitude and a factor of a few ($0.27-1.5 \times 10^{41}~{\rm erg~s^{-1}}$).
Significant \h{} features (e.g. Balmer lines) require two or three orders of magnitude more \h{} mass than predicted.

The production of elements between Li and K is negligible in all relevant kinds of ejecta. This very robust feature is due to the presence of $Z=N=20$ magic nuclear numbers, which prevents seeds nuclei formed in the iron group region to reach elements below calcium. Due to their low abundances, their observational impact in the early spectrum is also negligible.

Besides these very light elements, we investigated the production of strontium in our models, a light \rprocess{} element \citep[whose cosmological production is however dominated by the \sprocess, see e.g.][]{pra2020}, which has been possibly identified in the spectrum of AT2017gfo \citep{watson19}.
In agreement with previous findings \citep[see e.g.][]{wanajo2014,Radice:2018pdn,ned20}, we found that \sr{} is produced both in the dynamical and spiral wave wind ejecta.
In the former case, \sr{} production happens in the high-$Y_e$, high-latitude ejecta, and its amount ranges between 1 and 5 $\times 10^{-5} M_{\odot}$ (including our estimates of the uncertainties of numerical origin). In the latter case, a significant amount of \sr{} is synthesized at all angles due to the relatively large $Y_e$, with a possible exception of the region very close to the equator.
\sr{} production is very efficient and its typical mass fractions (integrated over the whole spiral wave wind ejecta) are between 1.5 and 4.5\%. The amount of \sr{} in the dynamical ejecta is comparable to the one required to explain the spectral features in AT2017gfo. Since the spiral-wave wind can efficiently produce even more \sr{}, our results suggest that the central object in GW170817 should have collapsed within $\sim 100~{\rm ms}$.

Since \h{} and \he{} are often observed in EM transients, our results could be relevant to organize and prioritize future observational campaigns for EM counterparts of GW events during the first days, when dynamical ejecta become transparent and kilonova spectra are still close to black body.
We remark that in this work we have limited our analysis to very light elements, up to potassium. However, it has been recently shown that calcium, for example, could provide a visible signature in kilonova spectra \citep{Domoto.etal:2021}. An extension of this study to elements heavier than potassium will be the subject of a future work.
Our results are robustly grounded on state of the art, first-principle simulations of BNS mergers and nucleosynthesis calculations. We have shown that the usage of a finite-temperature, composition dependent nuclear EOS is key to predict the correct amount of ejecta and its angular distribution. Moreover, the inclusion of weak interactions and neutrino irradiation are essential to correctly model the relative amount of neutrons and protons and, ultimately, to predict detailed nucleosynthesis yields to be compared with observations. Nevertheless, several limitations and weaknesses still affect our results. Our partial knowledge of the nuclear EOS and the still large uncertainties in the modeling of neutrino radiation, in addition to the lack of a robust numerical convergence of the ejecta properties, represent the largest limitation to our capability of predicting with accuracy the properties of the ejecta emerging from a BNS merger. Additionally, our limited knowledge of the relevant atomic opacities and the many uncertainties and approximations still present in the radiative transfer behind kilonova predictions still limit our predictive capacity. Our work underlined also the possible relevance of NLTE effects in shaping kilonova spectra. However, our approach mostly relied on supernova modeling, while NLTE effects in kilonovae have not been studied in details.
A strong effort in linking models to observations will be required to fill all these gaps in the years to come.

\acknowledgments
We thank the \texttt{ENGRAVE} collaboration and D. Malesani for useful discussions that have inspired this work and the \texttt{CoRe} collaboration for providing simulation data.
AP acknowledges the usage of computer resources under a CINECA-INFN agreement (allocation INF20\_teongrav). He also acknowledge PRACE for awarding him access to Joliot-Curie at GENCI@CEA.
AcF is partially supported by the PRIN-INAF 2017 with the project \textit{Towards the SKA and CTA era: discovery, localisation, and physics of transients sources} (P.I. M. Giroletti). 
SB acknowledges support by the EU H2020 under ERC Starting Grant, no.BinGraSp-714626. EC and MB acknowledge support from PRIN MIUR 2017 (grant 20179ZF5KS). 
DR acknowledges support from the U.S. Department of Energy, Office of Science, Division of Nuclear Physics under Award Number(s) DE-SC0021177 and from the National Science Foundation under Grant No. PHY-2011725. 
CV was supported for this work by the Excellence Cluster ORIGINS, which is funded by the Deutsche Forschungsgemeinschaft (DFG, German Research Foundation) under Germany's Excellence Strategy -- EXC-2094 -- 390783311.
AnF acknowledges support by the European Research Council (ERC) under the European Union’s Horizon 2020 research and innovation program (ERC Advanced Grant KILONOVA No. 885281)
NR simulations were performed on SuperMUC-LRZ (Gauss project pn56zo), Marconi-CINECA (ISCRA-B project HP10BMHFQQ and INF20\_teongrav allocation); Bridges, Comet, Stampede2 (NSF XSEDE allocation TG-PHY160025), NSF/NCSA Blue Waters (NSF  AWD-1811236), Joliot-Curie at GENCI@CEA (PRACE-ra5202) supercomputers. This research used resources of the National Energy Research Scientific Computing Center, a DOE Office of Science User Facility supported by the Office of Science of the U.S.~Department of Energy under Contract No.~DE-AC02-05CH11231.\\


\software{This research made use of \texttt{SkyNet} \citep{lipp17}, and of  \texttt{TARDIS}, a community-developed software
package for spectral synthesis in supernovae
\citep{tardis, kerzendorf_wolfgang_2019_2590539}.
The development of \texttt{TARDIS} received support from the Google Summer of Code initiative and from ESA's Summer of Code in Space program. \texttt{TARDIS} makes extensive use of Astropy and PyNE.}

\newpage

\appendix

\section{Origin of helium: fusion VS decay reactions}
\label{appendix: alpha}

\begin{figure*}
\centering
\includegraphics[width = \linewidth]{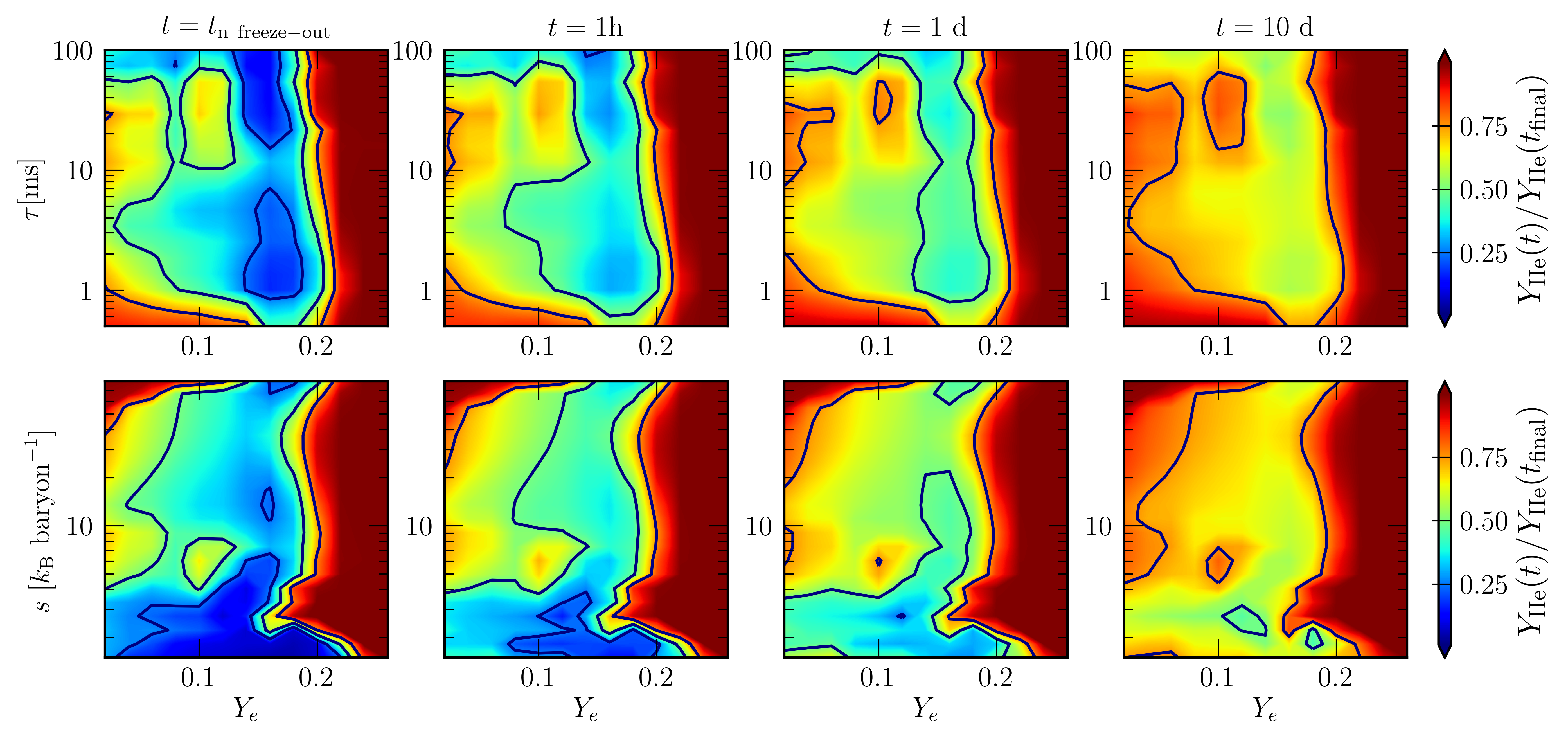} 
\caption{Ratio of the helium abundance produced at different times after merger over the final one. The ratios are represented in the $(Y_e,\tau)$ (top) and in the $(Y_e,s)$ (bottom) planes. Left panels, referring to the $n$ freeze-out time, represent the ratio of \he~produced through charged current reactions involving deuterium and tritium. The other three columns show the cumulative \he~production through $\alpha$-decays between 1 hour and 10 days. Blue lines refer to 0.25, 0.50 and 0.75.}
\label{fig:alpha origin}
\end{figure*}

In this appendix, we quantify the relative importance between charged reactions and $\alpha$-decays, together with its temporal evolution. To do that, we compare $Y_{\rm He}$ at the end of our calculations ($Y_{\rm He}(t_{\rm final})$) with $Y_{\rm He}$ at and after neutron freeze-out.
In figure \reffig{fig:alpha origin}, we show $Y_{\rm He}(t)/Y_{\rm He}(t_{\rm final})$ for different times (namely, the neutron freeze-out time, 1 h, 1 day and 10 days). In the top and bottom panels we consider subsets of the trajectories we have presented in \refsec{subsec:results nucleo} characterized by $s= 9.0~k_B~{\rm baryon^{-1}}$ and $\tau = 8.6 {\rm ms}$, respectively. We focus on $Y_e \lesssim 0.2$ since for larger $Y_e$ there are no $\alpha$-decaying nuclei and all \he{} is produced at $n$ freeze-out.
For $0.1 \lesssim Y_e \lesssim 0.2$, between $\sim$20 and $\sim$40\% of the final \he~is already produced at neutron freeze-out. For $Y_e < 0.1$ the relative amount tends to increase up to $\sim$70\%. Since many $\alpha$-decays happen on timescales of several days, $\lesssim$40\% ($\lesssim$25\%) of $Y_{\rm He}$ is produced after the first (10th) day after merger.

\section{Ion abundances in kilonova ejecta under LTE conditions}
\label{appendix: saha}

\begin{figure*}
\centering
\includegraphics[width=0.49\linewidth]{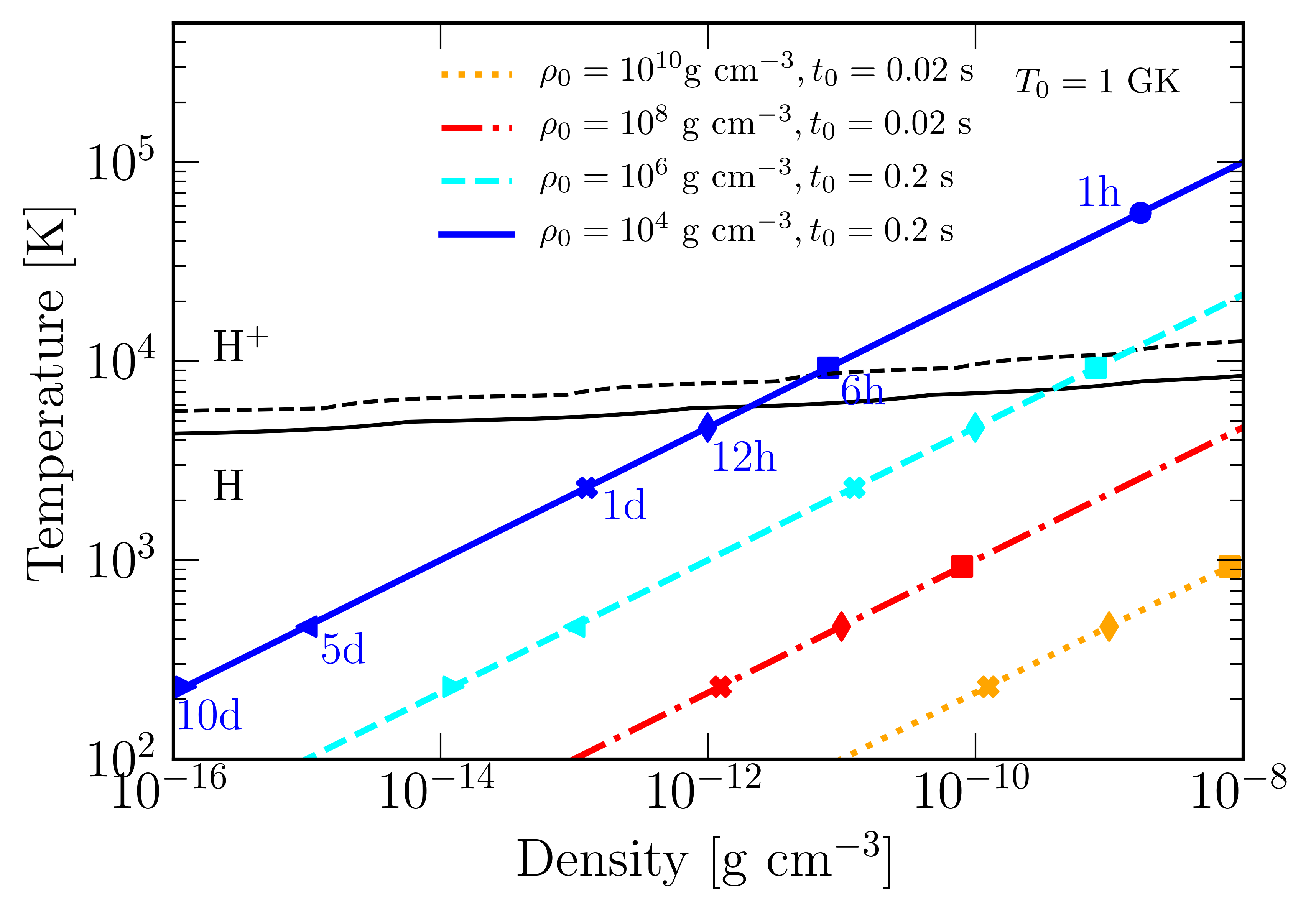}
\includegraphics[width = 0.49\linewidth]{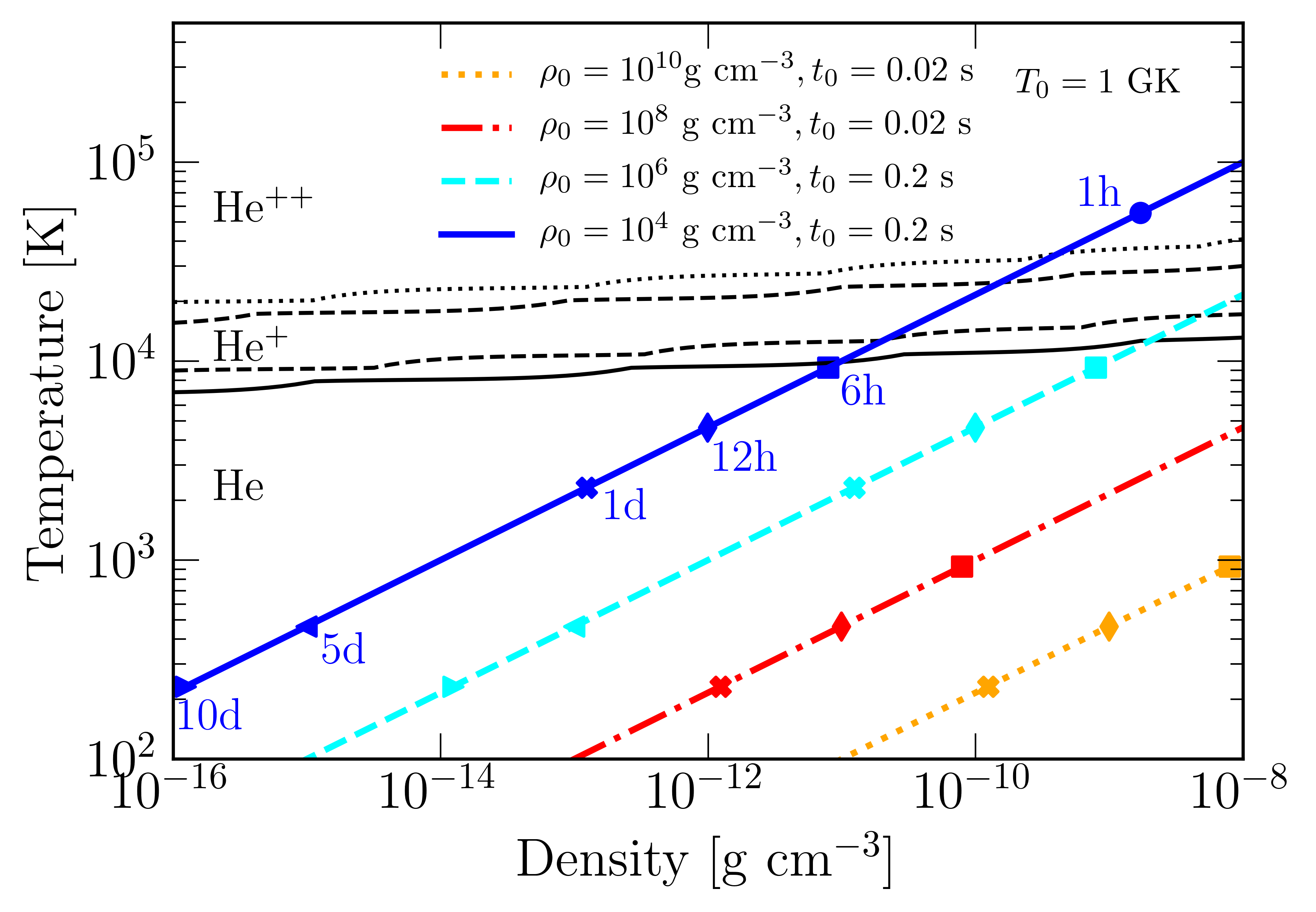}
\caption{Trajectories of expanding ejecta powering a kilonova, and \h{} (left) and \he{} (right) ion relative abundances in the $(\rho,T)$ plane under LTE conditions. Different lines correspond to different conditions: 
fast and slow expanding ejecta are characterized by $t_0 =0.02~{\rm s}$ and $t_0 =0.2~{\rm s}$, respectively. Different markers represent different times with respect to merger. 
To compute ion abundances, whole elemental abundances from representative tracers were assumed.
Left: dotted/dashed/solid black lines mark isocontours where the abundance of ${\rm He^{++}}$/${\rm He^{+}}$/atomic ${\rm He}$ (normalized to the total \he{} abundance) is 90\%. Thus, ${\rm He}^{++}$ dominates for $T \gtrsim 3 \times 10^4~{\rm K}$, while atomic \he{} for $T \lesssim 10^4~{\rm K}$. Right: dashed/solid black lines mark isocontours where the abundance of ${\rm H^{+}}$/atomic ${\rm H}$ (normalized to the total \h{} abundance) is 90\%. Thus, ${\rm H}^{+}$ dominates for $T \gtrsim  10^4~{\rm K}$.} 
\label{fig:ion abundances}
\end{figure*}

In this appendix, we present \h{} and \he{} ion abundances under LTE conditions by solving the Saha equation for a mixture of atoms and ions, as predicted by representative tracer particles.
In computing the atomic partition functions, we restricted ourselves only to the ground states, while for the ionisation energies and spins we considered the publicly available database provided by \cite{NIST_ASD}.
In practice, for \h{} we consider the full abundances produced by a tracer characterised by $\tau=1~{\rm ms}$, $s=100~{k_{\rm B}~{\rm baryon^{-1}}}$, and $Y_e = 0.15$, while for \he{} by $\tau=10~{\rm ms}$, $s=10~{k_{\rm B}~{\rm baryon^{-1}}}$, and $Y_e = 0.15$. We calculated the abundances of atomic and ionised \h{} and \he{} over a wide range of matter densities and temperature, namely: $ 10^{-16} \leq \rho [{\rm g~cm^{-3}}] \leq 10^{-8} $ and $ 100 \leq T~[{\rm K}]\leq 5 \times 10^5 $.
Doubly ionised helium dominates for $T \gtrsim 3 \times 10^4~{\rm K}$, while atomic \he{} for $T \lesssim 10^4~{\rm K}$. We tested the sensitivities of our results to the specific nuclear abundances by changing the entropy by one order of magnitude, without noticing qualitative differences. To explore conditions that are relevant for the kilonova emission, we consider profiles of density and temperature suitable to describe homologously expanding material: $\rho(t) = \rho_0 (t/t_0)^{-3}$ and $T(t) = T_0 (t/t_0)^{-1}$. While we keep $T_0=1~{\rm GK}$, we vary $\rho_0$ and $t_0$ to account for ejecta of different kinds. 
In particular, we consider $t_0 =0.02~{\rm s}$ and $\rho=10^{8}-10^{10}~{\rm g~cm^{-3}}$, and $t_0 =0.2~{\rm s}$ and $\rho=10^{4}-10^{6}~{\rm g~cm^{-3}}$, for fast and slow expanding ejecta, respectively. 
As shown in \reffig{fig:ion abundances}, for the density and temperature conditions expected during a kilonova and assuming LTE, \h{} and \he{} recombine to atomic form very rapidly. Only for slowly expanding ejecta, ionized states are present up to a few hours after the merger. 
In fact, while a few hours after explosions the ejecta temperature reaches $10^5$~K, already at 0.5 day the temperature drops to $10^4$~K. In these conditions, the small predicted masses of \h{} and \he{} are not expected to produce persistent lines, at least for LTE level populations.
These results depend very weakly on the specific trajectory.


\end{document}